\documentclass[usegraphicx,usenatbib]{mn2e}
\usepackage{graphicx}
\usepackage{epsfig}
\usepackage{amsmath}
\usepackage{mathrsfs}
\usepackage{amssymb}
\usepackage[english]{babel}
\usepackage{amsfonts}
\usepackage{fontenc}
\usepackage{times}
\usepackage{float}
\usepackage{txfonts}
\usepackage{textcomp}
\usepackage{booktabs}
\usepackage{microtype}
\usepackage{aas_macros}
\usepackage{xspace}
\usepackage[utf8]{inputenc}
\usepackage[T1]{fontenc}
\usepackage[usenames,dvipsnames,svgnames,table]{xcolor}
\usepackage[caption=false]{subfig}
\newcommand*{\rhoh}{\rho_{\rm h}}
\newcommand*{\rh}{r_{\rm h}}
\newcommand*{\rhocrit}{\rho_{\rm crit}}
\newcommand*{\deltac}{\delta_{\rm c}}
\newcommand*{\Mh}{M_{\rm h}}
\newcommand*{\rhoc}{\rho_{\rm c}}
\newcommand*{\rhostar}{\rho_*}
\newcommand*{\se}{\sigma_{\rm e8}}
\newcommand*{\Mstar}{M_*}
\newcommand*{\re}{R_{\rm e}}
\newcommand*{\Msun}{M_{\odot}}

\newcommand*{\fDM}{f_{\rm DM}}
\newcommand*{\alphaSN}{\alpha_{\rm SN}}
\newcommand*{\alphaWinds}{\alpha_*}
\newcommand*{\mSN}{\dot\rho_{\rm SN}}
\newcommand*{\mWinds}{\dot\rho_*}
\newcommand*{\Lsig}{L_{\sigma}}
\newcommand*{\Lstr}{L_{\rm v}}
\newcommand*{\Lrot}{L_{\rm rot}}
\newcommand*{\gth}{\gamma_{\rm{th}}}
\newcommand*{\Lx}{L_{\rm X}}
\newcommand*{\Tx}{T_{\rm X}}
\newcommand*{\Lkin}{L_{\rm{kin}}}
\newcommand*{\Tkin}{T_{\rm kin}}
\newcommand*{\Tsig}{T_{\sigma}}
\newcommand*{\Trot}{T_{\rm rot}}
\newcommand*{\Tphi}{T_\varphi}
\newcommand*{\tform}{t_{\rm form}}
\newcommand*{\tcool}{t_{\rm cool}}
\newcommand*{\theat}{t_{\rm heat}}
\newcommand*{\tdyn}{t_{\rm dyn}}
\newcommand*{\mh}{m_{\rm p}}
\newcommand*{\kb}{k_{\rm B}}

\newcommand*{\reo}{R_{\mathrm e\,0}}
\newcommand*{\Lb}{L_{\mathrm B}}
\newcommand*{\Lbs}{L_{\mathrm{B}\,\odot}}

\newcommand*{\Lm}{L_{\mathrm m}}

\newcommand*{\ergs}{\mathrm{erg~s}^{-1}}
\newcommand*{\kms}{\mathrm{km~s}^{-1}}

\newcommand*{\Lsn}{L_{\mathrm{SN}}}

\newcommand*{\gta}{\gamma_{\mathrm{th}}^{\mathrm \varphi}}

\newcommand*{\dV}{\,\mathrm{d}V}

\newcommand*{\vphi}{v_{\varphi}}

\newcommand*{\masstot}{M_\mathrm{{gas}}\xspace}
\newcommand*{\Mhot}{M_\mathrm{hot}\xspace}

\newcommand*{\Minj}{M_\mathrm{{inj}}\xspace}
\newcommand*{\Mesc}{M_\mathrm{{esc}}\xspace}

\newcommand*{\Lphi}{L_{\varphi}}
\newcommand*{\Tm}{T_\mathrm{m}\xspace}

\newcommand*{\tento}[1]{10^{#1}\xspace}
\newcommand*{\Tv}{T_{\rm v}\xspace}
\newcommand*{\dtpartial}[1]{\dfrac{\partial#1}{\partial t}}
\newcommand*{\emissivity}{\mathscr{L}}
\newcommand*{\emissivityX}{\varepsilon_\mathrm{X}}
\newcommand*{\uv}{\boldsymbol{u}}
\newcommand*{\vv}{\boldsymbol{v}}
\newcommand*{\uphi}{u_{\varphi}}
\newcommand*{\uR}{u_R}
\newcommand*{\uz}{u_z}
\newcommand*{\Rsn}{R_{\rm SN}}
\newcommand*{\gyr}{\mathrm{Gyr}}
\newcommand*{\myr}{\mathrm{Myr}}

\newcommand*{\kpc}{\mathrm{kpc}}
\newcommand*{\pc}{\mathrm{pc}}
\newcommand*{\mtot}{\dot\rho}
\newcommand*{\convective}[1]{\left(#1 \cdot \nabla \right)}
\newcommand*{\norm}[1]{\left\lVert#1 \right\rVert}

\newcommand*{\trace}{\mathrm{Tr}}
\newcommand*{\gradient}{\nabla}
\newcommand*{\diver}{\nabla\cdot}
\newcommand*{\veldisp}{\boldsymbol{\sigma}}
\newcommand*{\evphi}{{\bf e}_{\varphi}}
\newcommand*{\Sigmax}{\Sigma_\mathrm{X}}

\title[X-ray haloes and galaxy structure]
{The effects of galaxy shape and rotation on the X-ray haloes of
early-type galaxies - II. Numerical simulations}
 \author[A. Negri, S. Posacki, S. Pellegrini \& L. Ciotti]{Andrea Negri\thanks{E-mail:
andrea.negri@unibo.it}, Silvia Posacki\thanks{E-mail:
silvia.posacki@unibo.it}, Silvia Pellegrini \& Luca Ciotti
\\Department of Physics and Astronomy, University of Bologna, viale Berti Pichat
6/2, 40127 Bologna, Italy}

\date{2014 August 20}

\pagerange{\pageref{firstpage}--\pageref{lastpage}} \pubyear{2014}

\begin{document}
\maketitle
\label{firstpage}

\begin{abstract}
By means of high resolution 2D hydrodynamical simulations, we study
the evolution of the hot ISM for a large set of early-type galaxy 
models, characterized by various degrees of flattening and internal
rotation. The galaxies are described by state-of-the-art
axisymmetric two-component models, tailored to reproduce real
systems; the dark matter haloes follow the Navarro-Frenk-White or the
Einasto profile. The gas is produced by the evolving stars, and heated by Type Ia SNe.
We find that, in general, the rotation field of the ISM in rotating 
galaxies is very similar to that of the stars, with a consequent negligible
heating contribution from thermalization of the ordered motions.
The relative importance of flattening and rotation in determining the final X-ray
luminosity $\Lx$ and temperature $\Tx$ of the hot haloes is a function of the galactic mass.
Flattening and rotation in low mass galaxies favour
the establishment of global winds, with the consequent reduction of $\Lx$. In
medium-to-high mass galaxies, flattening and rotation are not sufficient to induce
global winds, however, in the rotating models the nature of the gas flows is deeply affected by conservation of
angular momentum, resulting in a reduction of both $\Lx$ and $\Tx$.
\end{abstract}

\begin{keywords}
galaxies: elliptical and lenticular, cD -- galaxies: ISM -- galaxies: kinematics and dynamics
-- X-rays: galaxies -- X-rays: ISM -- methods: numerical
\end{keywords}

\section{Introduction} 
\label{sec:intro}
Early-Type galaxies (ETG) are embedded in hot ($10^6 -10^7$~K), X-ray emitting
gaseous haloes \citep{fabbiano1989, o'sullivan2001}, produced mainly
by stellar winds and heated by Type Ia supernovae (SNIa) explosions
and by the thermalization of stellar motions. 
In particular, the thermalization of stellar motions is due to the interaction
between the stellar and SNIa ejecta and the pre-existing hot ISM (e.g., \citealt{Parriott.Bregman.2008}).
A number of different astrophysical phenomena determine the X-ray properties of the hot ISM,
such as stellar population evolution, galaxy structure and internal
kinematics, AGN presence, and environmental effects. In particular,
one of the empirical discoveries that followed the analysis of first X-ray data
of ETGs was the sensitivity of the hot gas content to major galaxy
properties as the shape of the mass distribution, and the mean
rotation velocity of the stellar component (see
\citealt{Kim.Pellegrini2012,mathews.brighenti.2003} for a full
discussion of the most relevant observational and theoretical aspects
concerning the X-ray haloes). From \textit{Einstein} observations it
was found that on average, at any fixed optical luminosity $\Lb$, rounder systems had
larger total X-ray luminosity $\Lx$ and $\Lx/\Lb$ (a measure of the galactic hot gas
content), than flatter ETGs and S0 galaxies \citep{eskridge.etal.1995}.
Moreover, galaxies with axial ratio close to unity spanned the full
range of $\Lx$, while flat systems had $\Lx\lesssim 10^{41}\ergs$.
This result was not produced by flat galaxies having a lower $\Lb$
with respect to round ETGs, since it held even in the range of $\Lb$
where the two shapes coexist \citep{pellegrini.1999}. Moreover it was
found that $\Lx/\Lb$ can be high only in slowly rotating galaxies, and
is limited to low values for fast rotating ones
\citep{pellegrini.etal.1997,sarzi.etal2010}. The relationship between $\Lx$ and
shape/rotation was reconsidered, confirming the above trends, for the
\textit{ROSAT} PSPC sample \citep{pellegrini2012}, and for the
\textit{Chandra} sample \citep{li.wang.etal2011,
boroson.etal2011,sarzi.etal.2013}. In particular, \citet{sarzi.etal.2013}, after
confirming that slow rotators generally have the largest $\Lx$ and
$\Lx/\Lb$ values, also found that their gas temperature $\Tx$ values are consistent just with the
thermalization of the stellar kinetic energy, estimated from $\sigma_{\mathrm e}$
(the luminosity averaged stellar velocity dispersion within the 
effective radius $\re$). Fast rotators, instead, have
generally lower $\Lx$ and $\Lx/\Lb$ values, and the more so the larger
their degree of rotational support; the $\Tx$ values of fast rotators
keep below 0.4 keV and do not scale with $\sigma _{\mathrm e}$.
Therefore, there seems to be a dependence of the hot gas
content and temperature on the galactic shape and internal dynamics. The
investigation of the origin of this still poorly understood
sensitivity is the goal of the present paper.

Theoretically, different possibilities have been proposed, and
explored both analytically (\citealt{ciottiPellegrini1996}, hereafter
CP96; \citealt{posacki.etal2013}, hereafter P13) and numerically 
(\citealt{brighenti.mathews.1996,dercole.ciotti1998}, hereafter DC98; \citealt{negri.etal2014}, hereafter N14).
The proposed explanations can be
classified in two broad categories: the {\it energetic} ones and the {\it 
hydrodynamical} ones. Of course, the difference is not sharp, as
the flow energetics affects the hydrodynamical evolution, and
different hydrodynamical configurations lead to different redistributions of the
energy available.

\textit{Explanations based on energetic effects} suppose that the
ISM in flat and rotating galaxies is less bound than in more round and
non rotating galaxies of similar luminosity (and so of similar SNIa energy
input), so that in the former objects the ISM is more prone to
develop a global/partial galactic wind, with the consequent decrease
of $\Lx$. A subdivision of energetic explanations is
represented by consideration of {\it flattening} effects vs. {\it
rotational} effects. The flattening explanation requires that flat
galaxies have shallower potential wells than rounder galaxies of similar
luminosity, so the gas is less bound (independently of the galaxy
kinematical support, ordered rotation or velocity dispersion).
In this scenario, the X-ray under-luminosity
of rotating galaxies is just a by-product of the fact that fast
rotation can be associated only to significant flattening. The rotational
explanation, instead, assumes that the gas injected in the galaxy
retains the stellar streaming motion, and so it is less
bound in rotating galaxies than in velocity dispersion supported galaxies of similar shape:
the correlation of X-ray under-luminosity
with galaxy flattening is now a by-product of the fact that rotating galaxies are generally flat.
Note that in this explanation the
thermalization of ordered ISM velocities is low, and then the ISM also lacks 
the corresponding energetic input; so the two effects (lower effective binding
energy but also lower heating) are in competition.
Indeed, the stellar random kinetic energy
is always supplied to the ISM, while the thermalization of the
ordered kinetic energy depends on the relative motion between the
stellar population and the ISM (see also \citealt{pellegrini2011}, P13).
P13 proposed that a lower energy injection at fixed $\Lb$, due to the lack of
thermalization of stellar streaming motions, could explain the lower
$\Tx$ of rotating galaxies. 
The energetic scenario has been explored analytically in CP96 by
using two-component Miyamoto-Nagai models, and in P13 by building a wide set of more realistic ETG models,
with different flattened structures and kinematics. 
These works showed that the binding energy of the gas depends on the procedure adopted to
``flatten'' the galaxy models, and that rotation affects also the hot gas temperature.
Preliminary numerical results for S0 galaxies suggest a low thermalization
in rotating systems (N14).

\textit{Explanations based on hydrodynamical effects} are less direct.
In this scenario the 
rotation of the gas injected in the galaxy leads to
hydrodynamical configurations of the ISM that, for different but cooperating
reasons associated with angular momentum conservation, are less X-ray
luminous than in non rotating systems.
For example, the gas density in the central galactic region is lower in rotating models,
where a rotationally supported cold disc forms, than in non-rotating ones, where the gas flows
straight to the centre, forming a hot, dense core. Thus,
the X-ray faintness is not due to the onset of galactic winds, but to
redistribution of the gas inside the galaxy. 
Now rotation is the main driver of X-ray under
luminosity, and correlation with galaxy flattening is a by-product.
Exploratory numerical simulations (N14) further showed that the X-ray under-luminosity of
flat objects could be due to hydrodynamical effects associated with angular
momentum conservation of gas injected at large radii for massive
galaxies, but it is more and more due to energetic reasons for low-mass systems.

Empirically, due to the difficulty to observationally disentangle
purely rotational and purely flattening effects, it is not easy to draw a clear cut
between the two broad explanations illustrated above. First of all,
since rotation is dependent on flattening, only galaxies sufficiently
flattened are expected to rotate significantly. Secondly, given that the
observed flattening in real objects is affected by projection effects,
intrinsically flattened galaxies, if observed face-on, can be
found in the region occupied by rounder galaxies in the
ellipticity-X-ray properties diagrams.

For these reasons, by using high-resolution 2D hydrodynamical simulations of gas flows in realistic,
state-of-the-art models of ETGs, we study this
long-standing issue of the X-ray under-luminosity of flat and rotating
galaxies, together with the properties of their $\Tx$ values. In order to derive robust
conclusions, we perform a large-scale
exploration of the parameter space of realistic (axisymmetric)
galaxy models characterized by different stellar mass, intrinsic flattening,
distribution of dark matter, and internal kinematics. In particular, the
galaxy flattening is supported by ordered rotation (isotropic
rotators) or by tangential anisotropy. All galaxy models are
constructed by using the Jeans code described in P13, and are tailored
to reproduce the observed properties and scaling laws of ETGs.
These simulations also allow to test how much simple energetic
estimates (such those of CP96 and P13) can be trustworthy in interpreting the global properties of
the hot gaseous X-ray coronae.

The paper is organized as follows. In Section 2 we describe the main
ingredients of the simulations, such as the galaxy models and the
input physics. The main results are presented in Section 3, and
Section 4 summarizes the conclusions.
\section{The simulations} 
\label{sec:sims}
\subsection{The galaxy models} 
\label{sec:galmod}
For the simulations we adopt axisymmetric two-component galaxy models,
where the stellar component can have different intrinsic flattening, while
for simplicity the DM halo is
kept spherical. In particular, the stellar
component is described by the deprojection \citep{mellier.etal1987} of
the \citet{devaucouleurs.1948} law, generalized for ellipsoidal
axisymmetric distributions,
\begin{equation}
\rho_*(R,z)=\rho_0\xi^{-0.855}\exp(-\xi^{1/4}),
\label{eq:rho_*}
\end{equation}
with
\begin{equation}
\rho_0=\dfrac{M_*b^{12}}{16\pi q\reo^3\Gamma(8.58)},\quad
\xi=\dfrac{b^4}{\reo}\sqrt{R^2+\dfrac{z^2}{q^2}},
\label{eq:rho_*2}
\end{equation}
where $(R,\varphi,z)$ are the cylindrical coordinates and $b\simeq7.67$.
The flattening is controlled by the parameter $q\leqslant 1$, so that the minor
axis is aligned with the $z$ axis.
$\reo$ is the projected half mass radius (effective radius) when the galaxy is
seen face-on;
for an edge-on view, the circularized effective radius is $\re=\reo\sqrt{q}$.
In the simulations we restrict to $q$ values of (1, 0.6, 0.3), corresponding
to E0, E4 and E7 galaxies
when seen edge-on.
For the DM halo we adopt the NFW
\citep{navarro.etal1997} or the \citet{einasto1965} profiles.

 \renewcommand\arraystretch{1.4}
\begin{table*}
\caption{Fundamental galaxy parameters for the NFW and Einasto sets of
models.}
 \begin{tabular}{ccccccccccc}
\toprule
 Name                             &$\Lb$          &$\re$ &$M_*$     &$\Mh$        &$\se^\mathrm{NFW}$   &$\se^\mathrm{EIN}$   &$\fDM^\mathrm{NFW} $ &$\fDM^\mathrm{EIN} $ &$c$      \\ 
                                  &$(10^{11}\Lbs)$&(kpc)&$(10^{11}\Msun)$&$(10^{11}\Msun)$&$(\kms)$&$(\kms)$       &        &        &        \\
  (1)                    & (2)           & (3)           & (4)    &   (5)             &   (6)          & (7)    & (8)    &   (9) & (10)  \\               
 \midrule                                                                       
 \midrule

E0$^{200}$                           &0.27           &4.09  &1.25             &25              & 200    &200     & 0.61   &   0.57 &   37 
   \\
 \midrule                                                                   
                                                                            
EO4$^{200}_{\rm{IS}}$                &0.27           &4.09  &1.25               &25              & 166    &166     & 0.63   &   0.59 &   37     
\\
EO4$^{200}_{\rm{VD}}$                &0.27           &4.09  &1.25               &25              & 179    &179     & 0.63   &   0.59 &   37     
\\
EO7$^{200}_{\rm{IS}}$                &0.27           &4.09  &1.25               &25              & 124    &124     & 0.66   &   0.62 &   37     
\\
EO7$^{200}_{\rm{VD}}$                &0.27           &4.09  &1.25               &25              & 148    &149     & 0.66   &   0.62 &   37     
\\
FO4$^{200}_{\rm{IS}}$                &0.27           &4.09  &1.25               &25              & 178    &179     & 0.59   &   0.55 &   37     
\\
FO4$^{200}_{\rm{VD}}$                &0.27           &4.09  &1.25               &25              & 191    &192     & 0.59   &   0.55 &   37     
\\
FO7$^{200}_{\rm{IS}}$                &0.27           &4.09  &1.25               &25              & 150    &151     & 0.57   &   0.53 &   37     
\\
FO7$^{200}_{\rm{VD}}$                &0.27           &4.09  &1.25               &25              & 178    &179     & 0.57   &   0.53 &   37     
\\
 \midrule                                                                   
 \midrule                                                                   
                                                                            
E0$^{250}$                           &0.65           &7.04  &3.35             &67              & 250    &250     & 0.59   &   0.55 &   28 
  \\
 \midrule                                                                   
                                                                            
EO4$^{250}_{\rm{IS}}$                &0.65           &7.04  &3.35               &67              & 207    &208     & 0.62   &   0.57 &   28     
\\
EO4$^{250}_{\rm{VD}}$                &0.65           &7.04  &3.35               &67              & 223    &224     & 0.62   &   0.57 &   28     
\\
EO7$^{250}_{\rm{IS}}$                &0.65           &7.04  &3.35               &67              & 154    &155     & 0.66   &   0.61 &   28     
\\
EO7$^{250}_{\rm{VD}}$                &0.65           &7.04  &3.35               &67              & 184    &185     & 0.66   &   0.61 &   28     
\\
FO4$^{250}_{\rm{IS}}$                &0.65           &7.04  &3.35               &67              & 223    &224     & 0.57   &   0.53 &   28     
\\
FO4$^{250}_{\rm{VD}}$                &0.65           &7.04  &3.35               &67              & 240    &241     & 0.57   &   0.53 &   28     
\\
FO7$^{250}_{\rm{IS}}$                &0.65           &7.04  &3.35               &67              & 189    &190     & 0.56   &   0.51 &   28     
\\
FO7$^{250}_{\rm{VD}}$                &0.65           &7.04  &3.35               &67              & 223    &224     & 0.56   &   0.51 &   28     
\\
 \midrule                                                                   
 \midrule                                                                   
                                                                            
E0$^{300}$                           &1.38           &11.79 &7.80             &160             & 300    &300     & 0.62   &   0.57 &   22 
   \\
 \midrule                                                                   
                                                                            
EO4$^{300}_{\rm{IS}}$                &1.38           &11.79 &7.80               &160             & 248    &249     & 0.64   &   0.60 &   22     
 \\
EO4$^{300}_{\rm{VD}}$                &1.38           &11.79 &7.80                &160             & 267    &269     & 0.64   &   0.60 &   22    
 \\
EO7$^{300}_{\rm{IS}}$                &1.38           &11.79 &7.80               &160             & 185    &185     & 0.68   &   0.64 &   22     
 \\
EO7$^{300}_{\rm{VD}}$                &1.38           &11.79 &7.80               &160             & 221    &223     & 0.68   &   0.64 &   22     
 \\
FO4$^{300}_{\rm{IS}}$                &1.38           &11.79 &7.80               &160             & 266    &268     & 0.60   &   0.55 &   22     
 \\
FO4$^{300}_{\rm{VD}}$                &1.38           &11.79 &7.80               &160             & 286    &288     & 0.60   &   0.55 &   22     
 \\ 
FO7$^{300}_{\rm{IS}}$                &1.38           &11.79 &7.80               &160             & 224    &225     & 0.59   &   0.54 &   22     
 \\
FO7$^{300}_{\rm{VD}}$                &1.38           &11.79 &7.80               &160             & 265    &267     & 0.59   &   0.54 &   22     
 \\
\bottomrule
\end{tabular}
\flushleft
Notes: $(1)$ Model name: E0 identifies the spherical progenitor, and the superscript is the value of $\se$. 
For the other models, the nomenclature is as follows: for example, 
FO4$^{200}_{\rm{IS}}$ means a face-on flattened E4 galaxy, obtained
from the E0$^{200}$ progenitor, with isotropic rotation.
$(2)$ Luminosities in the $B$ band.
$(3)$ Effective radius (for a FO view for FO-built models, and an EO view for
EO-built models). For FO-built models, the edge-on effective radius is reduced by a factor $\sqrt{q}$ (Sect.~\ref{sec:galmod}).
$(4)$ Total stellar mass. 
$(5)$ Total DM mass.
$(6)-(7)$ Stellar velocity dispersion, as the luminosity-weighted
average within a circular aperture of radius $\re/8$, for the NFW and
Einasto sets, respectively; for non-spherical models, $\se$ is the edge-on viewed value.
$(8)-(9)$ DM fraction enclosed within a sphere of radius $\re$ for
the NFW and Einasto sets, respectively.
$(10)$ Concentration parameter for the NFW set.
\label{tab:params}
\end{table*}
\renewcommand\arraystretch{1.}

All models belong to two different {\it sets}, defined by
the specific profile of the DM halo. The first set is characterized by
an untruncated NFW profile
\begin{equation} 
\rhoh (r)=\dfrac{\rhocrit~\deltac \rh} {r (1+r/\rh)^2},
\label{eq:NFW}
\end{equation}
where $\rhocrit=3H^2/8\pi G$ is the critical density for closure. We
refer to the DM mass enclosed within $r_{200}$ (the
radius of a sphere of mean interior density 200 $\rhocrit$), as to the
halo mass $\Mh$. Following \citet{navarro.etal1997}
\begin{equation}   
\deltac=\dfrac{200}{3} \dfrac{c^3}{\ln(1+c)-c/(1+c)},\quad c\equiv
\dfrac{r_{200}}{\rh}.
\end{equation}
The models in the second set are embedded in an Einasto profile
\begin{equation} 
\rhoh (r)=\rhoc e^{d_n-x},\quad x\equiv d_n \left(\dfrac{r}{\rh}\right)^{1/n}
\label{eq:Einasto}
\end{equation}
where $\rhoc$ is the density at the spatial half-mass radius $\rh$,
$n$ is a free parameter, and
\begin{equation}
d_n\simeq 3n-\dfrac{1}{3}+\dfrac{8}{1215 n}
\end{equation}
\citep{RetMon}.
The total gravitational field is computed by using the code described
in P13, and the same code is also used to solve numerically the Jeans
equations for the velocity fields of the stars, under the implicit
assumption of a two-integral distribution function. Ordered and
random motions in the azimuthal direction are described using the
\citet{satoh1980} $k$-decomposition.
Following P13, the model parameters are tuned to reproduce the
Faber--Jackson and the size--luminosity relations as given by
\citet{desroches.2007} for $\approx 10^5$ ETGs in the
SDSS; the stellar mass-to-light ratios adopted pertain to a 12 Gyr old
stellar population with a Kroupa initial mass function \citep{maraston.2005}.

In each of the two sets, we consider different {\it families} of
models, built following the procedure described in P13 (Sects.~3.3 and 3.4).
Here we just recall the main steps. Each family is associated
with a spherical galaxy, that we call the ``progenitor''. The progenitor
structural parameters are determined by assigning $\se$ (the aperture
luminosity-weighted velocity dispersion within $\re/8$), and then
deriving the luminosity and effective radius $\re$ of the
galaxy from the scaling laws cited above. Then, from a chosen stellar
mass-to-light ratio, the stellar mass $\Mstar$ is derived.
Finally, the parameters of the DM halo are determined in order to reproduce the assumed
$\se$ and fixing $\Mh/\Mstar\simeq 20$ \citep{behroozi.etal2013}.
In the NFW set, these constraints produce $\rh\simeq 2\re$, $22 \lesssim c \lesssim 37$, 
and a DM fraction $\fDM$ within a
sphere of radius $\re$ of $\simeq 0.6$ for the spherical progenitors.
For the Einasto set we fix $n=6$, and we find that $\rh\simeq 20\re$, and
$\fDM\simeq 0.56$
for the spherical progenitor. 

In each of the two sets we considered three values of $\se$ for the
spherical progenitors, i.e., 200, 250 and 300 $\kms$. Therefore, each of the
two sets is made of 3 families of models, for a total of 6 spherical progenitors.
Table~1 lists all the relevant parameters characterizing the progenitors galaxy
models for both sets.
The flattened descendants of each progenitor with intrinsic
flattening of E4 ($q=0.6$) and E7 ($q=0.3$), are derived as follows.
We produce two flattened models for each value of $q$.
The first flattened model is called ``face-on built'' (hereafter FO-built), since, when observed face-on, its $\re$ is the same
as that of the spherical progenitor;
this requires FO-built flattened models to be more and more concentrated as $q$ decreases ($\rhostar\propto q^{-1}$). 
The second flattened model instead, when seen edge-on, has the same circularized $\re$ of the spherical progenitor, thus we call it ``edge-on built'' (hereafter EO-built);
this property makes the EO-built models expand with decreasing $q$ ($\rhostar\propto \sqrt{q}$).
Therefore, a spherical progenitor with a given value of $\se$ produces four flat galaxies: two E4 models (FO
and EO built), and two E7 models (FO and EO built). As a further step,
in order to study the effects of galaxy rotation, we assume two
kinematical supports for each flattened system: one corresponding to a
velocity dispersion supported galaxy (VD models), and the other one to
an isotropic rotator (IS models). These two configurations are
obtained by setting the Satoh parameter $k$ equal to 0 and 1,
respectively (see P13, eqs.~B3 and B4). In the flattening procedure the DM halo
is maintained fixed to that of the progenitor.
Note that our flattened models are representative of ETGs since they are consistent with their observed properties.
We indeed checked for models lying outside the observed scatter of the scaling laws, but our adopted flattening
procedure is quite robust in producing acceptable models, so that we retained all of them.

Summarizing, from each spherical progenitor of given $\se$, eight
flattened models are obtained (see Tab.~1), and we refer to this group of nine galaxy
models as to a family. All models belonging to a family can be identified
either by the $\se$ value of the spherical progenitor, or by their
stellar mass $\Mstar$ (or B luminosity), or DM halo mass;
note however that while these last three quantities are kept constant within a
family, the $\se$ of the descendants varies. Indeed, the modification
of stellar structure involves a change in the stellar kinematics, and
so in the value of $\se$; in particular, for our models $\se$
decreases for increasing flattening (see P13 for a comprehensive
discussion). Note that $\se$ depends on the line-of-sight direction
for non-spherical models; when quoting $\se$ for the latter models, 
in the following, we refer to the edge-on projection.

\begin{figure*}
\includegraphics[width=0.66\linewidth, keepaspectratio]{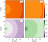}
\caption{Meridional section of temperature (in K, top panels) and heating over cooling time ratio $\theat / \tcool$
(bottom panels), for the E0$^{250}$ model with NFW halo (Table 1), at the times specified in the boxes (in Gyr).
We define $\tcool = E/\emissivity$ and $\theat$ as the ratio between $E$ and the 
source terms in the r.h.s of Eq.~(9). In the bottom plots, green regions refer
to cooling gas, while purple indicate heating dominated
regions, as indicated by the colour scale. Arrows show the meridional velocity field, with
the longest arrows corresponding to $127~\kms$.}
\label{fig1}
\end{figure*}
\subsection{Hydrodynamical equations} 
\label{sec:inpphys}

The input physics is fully described in DC98 and N14; here we
just summarize the main ingredients. We account for
mass sources due to stellar mass losses and SNIa ejecta, momentum
injections due to the azimuthal streaming motion of the stellar
population, and energy sources due to the thermalization of stellar motions and SNIa explosions.
Star formation and feedback from a central black hole are not
considered. The corresponding equations are
\begin{gather}
\dtpartial{\rho} + \diver (\rho\uv) = \mSN + \mWinds
\equiv \mtot 
\label{eq:continuity},\\
\rho\dtpartial{\uv} + \rho
\convective{\uv}\uv = - \gradient p - \rho
\gradient \Phi_{\rm tot} + \mtot\,(\vv - \uv) 
\label{eq:euler},\\
\begin{split}
\dtpartial{E} + \diver(E \uv) &= 
- p\diver \uv - \emissivity + \mSN \dfrac{u_s^2}{2} \\
&+ \dfrac{\mtot}{2} \left[ \norm{\vv - \uv}^2 +
  \trace (\veldisp^2) \right],
\label{eq:energy}
\end{split}
\end{gather}
where $\rho$, $\uv$, $E$, $p$, $\Phi_{\rm tot}$, $\emissivity$, $u_s$,
$\vv=\vphi\evphi$ and $\veldisp^2$ are respectively the ISM mass density,
velocity, internal energy density, pressure, total gravitational
potential, bolometric radiative losses per unit time and volume, the
velocity of the SNIa ejecta ($u_s \simeq 8.5 \times 10^3 ~ \kms$, corresponding to
$10^{51}~\mathrm{erg}$ associated with an ejecta of 1.4~$\Msun$), the
streaming velocity and the velocity dispersion tensor of the stellar
component. The ISM is considered as a fully ionized monoatomic
gas, with $p = (\gamma -1)E$, $\gamma = 5/3$, and $\mu\simeq 0.62$; as
usual the gas self-gravity is neglected.
 
The total mass injection rate per unit volume $\mtot$ has a
contribution from SNIa events $\mSN=\alphaSN(t) \rhostar$ and from
stellar winds $\mWinds =\alphaWinds(t) \rhostar$, where
\begin{equation}
\alphaSN (t) = \dfrac{1.4 \Msun}{\Mstar}\Rsn (t),\quad
\alphaWinds (t) = 3.3 \times 10^{-12} t_{12}^{-1.3} ~(\mathrm{yr^{-1}}).
\label{eq:SNIa}
\end{equation}
The SNIa explosion rate $\Rsn (t)$ is given by 
\begin{equation}
\Rsn (t) = 0.16\, h^2 \times 10^{-12}\Lb\, t_{12}^{-s} ~
(\mathrm{yr^{-1}}), 
\label{eq:SNIarate}
\end{equation}
where $h=H_0/70\,\mathrm{km~s^{-1}~Mpc^{-1}}$, $\Lb$ is the present epoch B-band
galaxy luminosity in blue solar luminosities, $t_{12}$ is the age of
the stellar population in units of 12 Gyr, and $s$ parametrizes the
past evolution. The SNIa's heating rate $\Lsn(t)$ is obtained from
Eq.~(11) by assuming that each SNIa event releases $10^{51}$~erg, and
we adopt $s=1$ in Eq.~(11).
Since the volume integrated energy and mass injections are dominated
by $\Lsn$ and stellar mass losses, respectively, this choice of $s$
produces a secular time-increase of the specific heating $\Lsn /\mtot$
of the input mass, due to the different time dependence of the mass
and energy inputs \citep[e.g., ][]{pellegrini2012}.

The radiative cooling is implemented by adopting a modified version of
the cooling law reported in \citet{sazonov2005}, with the lower limit
for the ISM temperature of $T=10^4$~K (see N14 for more details).
We also calculate the X-ray emission in the 0.3--8
keV {\it Chandra} band, and the X-ray emission weighted
temperature as
\begin{equation}
\Lx = \int \emissivityX dV, \quad \Tx = \dfrac{\int T \emissivityX dV}{\Lx},
\label{eq:Tx} 
\end{equation}
respectively, where $\emissivityX$ is the thermal emissivity in the
energy range 0.3--8 keV of a hot, collisionally ionized plasma,
obtained by the spectral fitting package
\textsc{xspec}\footnote{http://heasarc.nasa.gov/xanadu/xspec/.}
\citep[spectral model \textsc{apec},][]{smithetal2001}, 
and the volume integrals are performed over the whole computational mesh.

As in N14 all the simulations are performed with the ZEUS-MP2 code in
cylindrical coordinates, in order to better and uniformly resolve the galaxy
equatorial region. The adopted 2D axisymmetric $(R,z)$ grid is
organized in $480 \times 960$ logarithmically spaced points, extending
out to $\simeq 100$ kpc, with a resolution of $\simeq 90~\pc$ in the
central $10~\kpc$. We assume that at the beginning of the simulation
each galaxy is 2 Gyr old and is depleted of gas due to the feedback of
the stellar population; the evolution is followed for 11 Gyr, until an
age of 13 Gyr.

\subsection{The contribution of stellar kinematics to the ISM energetics}

The main issue investigated in this paper is the
effect of flattening and ordered rotation on the X ray luminosity
and temperature of the ISM. Analytical studies, based on the global
energetics arguments, showed that different and competitive effects should be
taken into account (CP96, \citealt{pellegrini2011}, P13).
Some of the expectations have been confirmed by past numerical studies, even though the
galaxy models adopted were not tailored on realistic elliptical
galaxies, but more on S0/Sa (DC98, N14). In any case, such studies
showed that different physical phenomena can be important
as a function of the galaxy mass and potential well depth (i.e.,
whether the ISM is outflowing or inflowing).

In order to compare the results of the numerical simulations with the global energetic estimates,
the following quantities are also computed by the hydrodynamical code.
The first is the thermalization
of stellar random motions, providing an energy input per
unit time to the ISM of
\begin{equation}
 \Lsig\equiv \dfrac{1}{2}\int \mtot\, \trace (\veldisp^2) \dV.
\label{eq:Lsigma}
\end{equation}
Note that while the contribution from stellar random motions is 
fully independent of the ISM velocity field (see Eq.~(8)), the 
thermalization of the stellar ordered (streaming) motion 
depends on the relative motion between stars ($\vv$) and ISM ($\uv$)
\begin{equation}
 \begin{split}
\Lstr & \equiv \dfrac{1}{2}\int \mtot \norm{\vv -\uv}^2 \dV \\
 & = \dfrac{1}{2}\int \mtot (\uR^2 +\uz^2) \dV + \dfrac{1}{2}  \int \mtot (\vphi
-\uphi)^2 \dV  =  \Lm+\Lphi,
\end{split}
\label{eq:Lstr}
\end{equation}
so that, at variance with $\Lsig$, it cannot
be predicted a priori from the knowledge of the galaxy structure and
kinematics. Note that in our galaxy models $\vv =\vphi\evphi$,
and $\Lm$ and $\Lphi$ are respectively the energy input rate due to the ISM
velocity in the meridional plane $(R,z)$, and to the \textit{relative} velocity of
stars and the ISM in the azimuthal direction.

As in P13 and N14, we parametrize the thermalized fraction of the available
kinetic energy due to stellar streaming with
\begin{equation}
\gth \equiv \dfrac{\Lstr}{\Lrot},
\label{eq:gth}
\end{equation}
where
\begin{equation}
\Lrot\equiv \dfrac{1}{2}\int \mtot\, \norm{\vv}^2  \dV 
\end{equation}
is the energy input per unit time that would be injected in a galaxy
with an ISM at rest (i.e., $\uv=0$), due to thermalization stellar
streaming motions. 
Note that $\gth$ is undefined (formally, it
diverges) for VD supported models, and can be very large for slow rotators and/or for gas
flows with large velocities in the meridional plane (as in the case of
galactic winds).
Using these definitions, the total energy supplied
to the ISM due to stellar motions can be written as
\begin{equation}
 \Lkin \equiv \Lsig + \Lstr = \Lsig + \gth \Lrot.
 \label{eq:Lkin}
\end{equation}

Finally, all the luminosities defined above can be converted into
equivalent temperatures as
\begin{equation}
\Tsig=\dfrac{\mu\mh}{3\kb\Mstar} \int\rhostar\mathrm{Tr}(\veldisp^2)\dV,\quad \Trot=\dfrac{\mu\mh}{3\kb\Mstar}\int\rhostar\norm{\vv}^2\dV,
\label{eq:tsig}
\end{equation}
where $\kb$ is the Boltzmann constant and $\mh$ is the proton mass, so that
\begin{equation}
\Tkin= \Tsig+\gth\Trot ;\quad \gth \Trot = \Tm + \Tphi.
\label{eq:tstar}
\end{equation}
\section{Results} 
\label{sec:results}
Here we present the main results of the simulations, focussing on the
hydrodynamical evolution of a few representative models, and then on the
global properties $\Lx$ and $\Tx$ for the two sets of models.
\subsection{Hydrodynamics} 
\label{sec:hydro}
For illustrative purpose, we present the hydrodynamical evolution of some
selected EO-built models belonging to the NFW set. 
In particular, in the family derived from the E0 progenitor with $\se =
250~\kms$ (Sect.~3.1.1), 
we consider the velocity dispersion supported E7 model (Sect.~3.1.2), and the
corresponding E7 isotropic rotator
(Sect.~3.1.3). In Sect.~3.1.4 we summarize the main
similarities and differences with the other models, as well as some
considerations on the behaviour of the thermalization parameter $\gth$.

In general, as found in N14, the gas flows are found to evolve
through two well defined
hydrodynamical phases. At the beginning, all the ISM quantities (density,
internal energy and velocity) are nearly symmetric with respect to
the galactic equatorial plane. During the evolution, the velocity
fields become more and more structured, until, after a certain 
time that depends on the specific model, the reflection symmetry is
lost, and it is never restored.
\begin{figure}
\includegraphics[width=\linewidth, keepaspectratio]{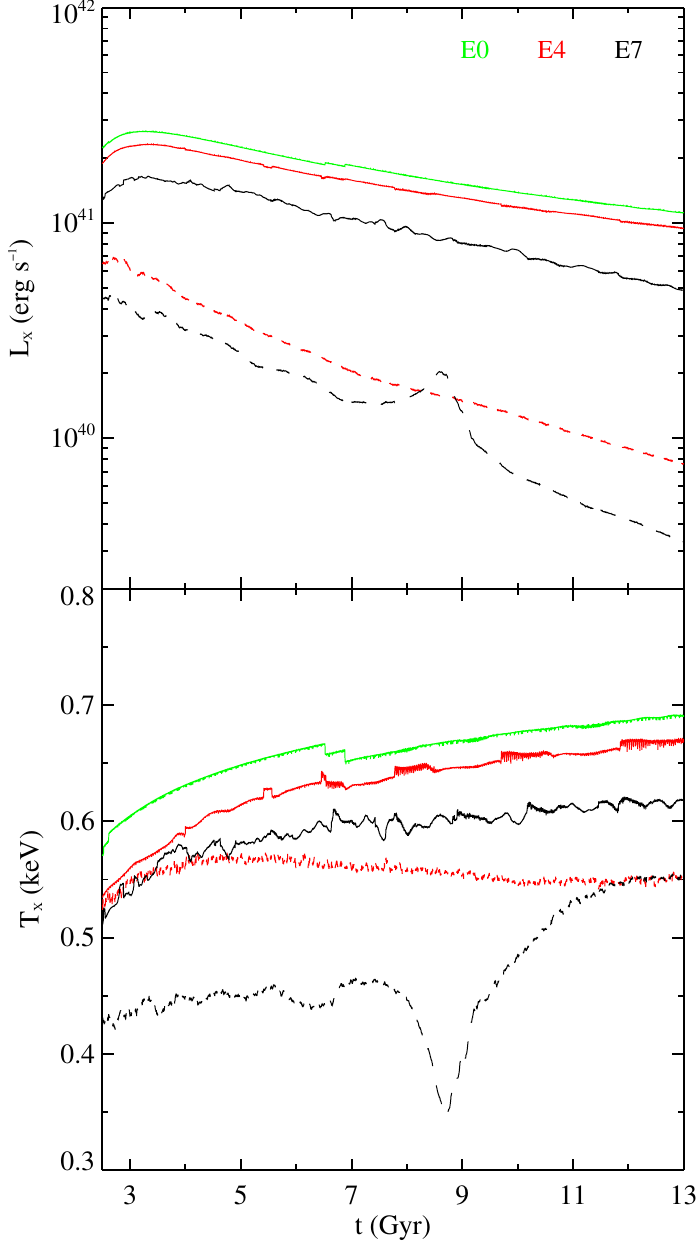}
\caption{Time evolution of the X-ray luminosity $\Lx$ and X-ray
emission weighted temperature $\Tx$ for the family derived from the E0$^{250}$ model with
the NFW halo. The red and
black lines report the
evolution of the VD models (solid), and of the IS
models (dashed). The colours map the flattening: green, red and
black correspond to the E0, E4 and E7 galaxies,
respectively.} \label{fig2}
\end{figure}

\subsubsection{The E0$^{250}$ progenitor}

The initial ($t=2.4~\gyr$) and final ($t=13~\gyr$) configurations of the ISM 
are shown in Fig.~\ref{fig1}, where we show the meridional section of the ISM temperature (top panels),
and the ratio of the heating and cooling time $\theat/\tcool$ (bottom panels; green corresponds to a cooling
dominated region while violet refers to a heating region). The arrows show the meridional velocity field.

All the ISM physical quantities are stratified
on a spherical
shape, as a consequence of the galaxy spherical symmetry. 
A decoupled flow is soon established ($t\simeq 2.4~\gyr$),
with an inflow in a round central region surrounded
by an outflowing atmosphere. At the same time cold gas
accumulates into the centre, due to the lack of rotational support.
Starting from the time of decoupling, the evolution appears to be nearly stationary.

The evolution of the ISM temperature reflects the flow evolution: an
hot atmosphere approximately isothermal ($T\simeq 5\times\tento{6}$ K) forms at the
beginning, containing a cooling region of radius $\simeq 5$ kpc that leads to the formation
of a cold core at the very centre (see the green region in the bottom panels).
At the end of the simulation, a total of $\simeq 2.6\times\tento{10} \Msun$ of gas are cooled at the centre,
while $\simeq 5 \times\tento{9} \Msun$ have been ejected as a galactic outflow.
Overall, $\Lx$ and $\Tx$ of this model do not present significant
fluctuations (Fig.~\ref{fig2}, solid green line), with $\Lx$ steadily decreasing and $\Tx$ steadily
increasing in pace with the time evolution of mass sources and specific heating.

\begin{figure*}
\includegraphics[width=0.66\linewidth, keepaspectratio]{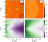}
\caption{Meridional sections of the temperature (top panels) and
 heating over cooling time ratio (bottom panels) for the EO$7^{250}_{\mathrm{VD}}$ 
model of the NFW set, at the times specified in the boxes (in Gyr).
Arrows are normalized to the same velocity as in Fig.~\ref{fig1}.}
\label{fig3}
\end{figure*}

\subsubsection{The EO$7^{250}_\mathrm{VD}$ galaxy}

The ISM evolution of the velocity dispersion supported EO$7^{250}_\mathrm{VD}$ model
presents important similarities with the spherical progenitor. This is not
surprising, due to the absence of angular momentum, and to the fact
that in general the gravitational potential is much rounder than the associated
stellar
density distribution (in addition, recall that the DM halo is kept spherical).
Therefore, the only major differences between the E0$^{250}$ progenitor and the
EO$7^{250}_\mathrm{VD}$ model are the different spatial regions where the gas
is injected, and the different velocity dispersion field of the stars.
A direct comparison of the evolution of the two models can be obtained by inspection of Fig.~\ref{fig3}, 
analogous of Fig.~\ref{fig1}. At early times the flow is
kinematically decoupled, with an equatorial outflow due to the
concentrated heating on the equatorial plane (purple region), associated with a
polar
accretion along the $z$-axis, evidenced by the green cooling region.
As in the spherical progenitor, due to the lack of
centrifugal support, the cooling material falls directly towards the
galaxy centre, where a dense, cold core is formed during the first
hundred of $\myr$. The early flow evolution is characterised by equatorial
symmetry and large scale meridional vortexes. The symmetry is lost at
$t \simeq 3$ Gyr, followed by a secular decrease of the velocity
field. The flow velocities are larger that in the E0 progenitor, due to 
the weaker gravitational field (consequence of the edge-on flattening).
The evolution of $\Lx$ and $\Tx$ are shown in Fig.~\ref{fig2} (black solid line).
Compared to the E0 progenitor, the EO$7^{250}_\mathrm{VD}$ model has a 
fainter $\Lx$ and a lower $\Tx$, but a similar lack of significant 
fluctuations in $\Lx$ and $\Tx$.
At the end of the simulation, the cooled gas at the centre for this model is $\simeq
1.8\times\tento{10} \Msun$, 
while the ejected ISM is $\simeq 7 \times \tento{9} \Msun$.
If this model is allowed to have the accretion physics activated for the central
black hole, we expect to recover the complex AGN feedback phenomena described elsewhere
\citep{Ciotti.Ostriker.2012},
with significant reduction of the central accreted mass.
%
\begin{figure*}
\includegraphics[width=\linewidth, keepaspectratio]{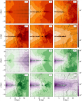}
\caption{Meridional sections of the temperature in K (six top panels) and 
heating over cooling time ratio (six bottom panels) for the EO$7^{250}_{\mathrm{IS}}$ model
 of the NFW set, at the times specified in the boxes (in Gyr). Arrows
are normalized to the same velocity as in Fig.~\ref{fig1}.}
\label{fig4}
\end{figure*}

\subsubsection{The EO$7^{250}_\mathrm{IS}$ galaxy}
Previous explorations (DC98, N14) revealed that the
evolution of gas flows in galaxies with significant ordered rotation of the
stellar component 
is more complex than in velocity dispersion supported systems of similar
structure. This is confirmed by the present study. The flow evolution of the EO$7^{250}_\mathrm{IS}$ is shown in 
Fig.~\ref{fig4} (where more panels than the previous two models are shown, to better illustrate the more structured
evolution of the ISM).

The first major difference of the present model with respect to its VD counterpart is the
formation, due to angular momentum conservation, of a rotationally
supported, thin and dense cold disc, with a size of $\simeq 5~\kpc$.
The cold disc grows during
galaxy evolution, reaching a final size of $\simeq 10~\kpc$.
A hot and rarefied zone that secularly increases in size surrounds the cold disc.
At early times ($t\simeq 2.1$ Gyr), the ISM in the central regions cools
and collapses, producing a low-density region that cannot be replenished
by the inflowing gas, which is supported by angular momentum.
As time increases, the combination of the centrifugal barrier, that
keeps the centre at low density, and the secular increase of the
specific heating produce the growth of the heating region
(purple zone in Fig.~\ref{fig4}, roughly extending as the cold thin disc).
We stress that the time and spatial evolution of the $\theat/\tcool$ ratio 
is more affected by cooling time variations than by the secular decrease of the heating
time. Being the cooling time very sensitive to the ISM density,
$\theat/\tcool$ is strongly related to the density distribution evolution. 

Another important difference between EO$7^{250}_\mathrm{IS}$ and EO$7^{250}_\mathrm{VD}$
concerns the ISM kinematics outside the equatorial plane. As apparent in Fig.~\ref{fig4}, starting from $t\simeq 8~\gyr$ the 
meridional velocity field develops a very complex pattern of
vortexes above and below the equatorial plane. 
This behaviour is associated with the formation of a large cooling region 
(in green), and corresponds respectively to a peak and a drop in the evolution 
of $\Lx$ and $\Tx$ (Fig.~\ref{fig2}, black dashed line).
Note also that $\Lx$ and $\Tx$ are the lowest of the three models E0$^{250}$, 
EO$7^{250}_\mathrm{VD}$ and EO$7^{250}_\mathrm{IS}$. The cold mass accreted at
the centre is now much smaller ($\simeq 2.9 \times \tento{3} \Msun$), while the mass in the cold disc is $\simeq
1.5 \times \tento{10} \Msun$,
and the mass ejected in the galactic wind is $\simeq 1.1\times\tento{10} \Msun$.
Note that a central black hole in this rotating model would produce
a significantly weaker AGN activity than in the EO$7^{250}_\mathrm{VD}$ model.

Even if the models are structurally quite different from the S0/Sa models
in N14, they show a similar ISM evolution, with a lower $\Lx$ and $\Tx$ in
rotating models. However, the number of oscillations in the present set of
models is much lower.
\begin{figure*}
\includegraphics[width=\linewidth, keepaspectratio]{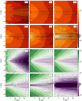}
\caption{Meridional sections of the temperature in K (six top panels) and
heating over cooling time ratio (six bottom panels) for the low mass EO$7^{200}_{\mathrm{VD}}$ model
of the NFW set, at the times specified in the boxes (in Gyr). Arrows
are normalized to the same velocity ad Fig.~\ref{fig1}.
Note the strong equatorial degassing established at late times.}
\label{fig5}
\end{figure*}
\subsubsection{The thermalization parameter and an overview of all models}
\begin{figure*}
\includegraphics[width=0.8\linewidth, keepaspectratio]{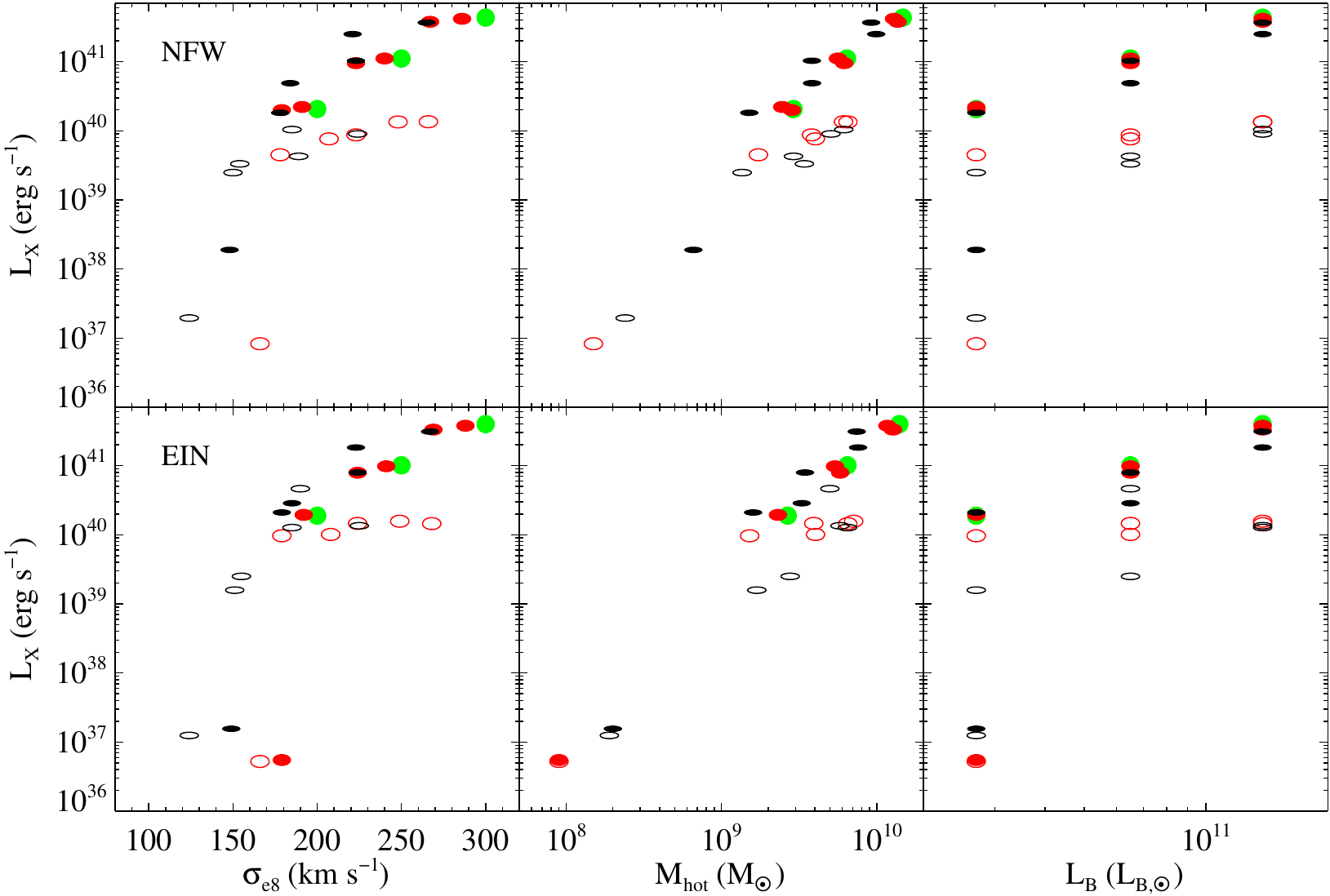}
\caption{ISM X-ray luminosity $\Lx$ in the 0.3--8 keV band at 13 Gyr
for all models in the NFW (top panels) and in the Einasto
(bottom panels) sets, as a function of $\se$, of the hot
($T> \tento{6}$~K) ISM mass, and of the galaxy blue optical luminosity;
spherical progenitors (green circles) with
$\se=(200, 250, 300)$ have been considered. The
green, red and black colours refer to the E0, E4 and E7 models
respectively. Filled and empty symbols indicate the
fully velocity dispersion supported VD models, 
and the isotropic rotators IS models, respectively.}
\label{fig6}
\end{figure*}
A useful global parameter that helps to quantify the heating of the ISM
due to stellar
ordered motions is the thermalization parameter $\gth$ (Sec.~2.3, P13, N14). 
For the EO$7^{250}_\mathrm{IS}$ galaxy we found that $\gth$ 
remains at low values ($\simeq 0.08-0.28$, see Tables~\ref{tab:NFW} and~\ref{tab:Ein}) over the evolution; this 
means that 1) the ISM almost co-rotates with the stellar population everywhere, and 
2) there are not significant ISM
velocities in the meridional plane. Note that $\gth$ can attain large values,
even larger than unity (see the examples described later in this
Section). This happens in general in low-rotation, low-mass systems,
where $\gth$ is fully dominated by high velocity galactic winds (see Eq.~14),
so that $\Lm$ is large even though $\Lphi$ remains low.
Remarkably enough, also in these high-$\gth$ cases, the azimuthal
thermalization parameter $\gta \equiv \Lphi/\Lrot$ (Eq.~14) remains low (see Tables~\ref{tab:NFW} and~\ref{tab:Ein}),
indicating that the ISM rotates almost as fast as the stellar
population. One could be tempted to interpret the lack of thermalization of a significant fraction of ordered
motions in all the IS models as the reason for
the lower $\Tx$ of the IS models with respect to their VD counterparts (Fig.~\ref{fig2},
black and red dashed lines vs. the solid lines). However, even if this effect
certainly contributes, it is not the main reason for the lower $\Tx$ in
rotating models. Indeed, we found that artificially adding the 
``missing'' thermalization to the equations of hydrodynamics in dedicated test simulations of
rotating models leads only to a negligible increase in $\Tx$ (see also N14), showing that
also other effects contribute to the low $\Tx$ (see Sect.~3.3).

We now discuss similarities and differences of the hydrodynamical
evolution in galaxy models of different mass (i.e., derived
from progenitors with different $\se$).
The main features of the family with the spherical progenitor of $\se =
250~\kms$ are maintained in the $\se = 300~\kms$ family. In particular,
independently of the DM halo profile, increasing $\se$, both $\Lx$ and $\Tx$ increase.
This is expected because more massive models can retain more and hotter gas independently of the flattening and
kinematical support. In more massive models the $\Lx$ and $\Tx$ are less
fluctuating with time, the outflow velocities of the galaxy outskirts are
lower, and the complicated meridional circulation in the rotating models is
reduced (as also found by N14).
The final properties of all models are given in Tables~\ref{tab:NFW}
and \ref{tab:Ein}. A full
discussion of $\Lx$ and $\Tx$ is given in Sects.~3.2 and 3.3.
In general, the ISM temperature, luminosity, radius of the central cooling
region, and inflow velocity are directly proportional to $\se$. In massive 
($\se \geq 250~\kms$) models, at fixed galaxy mass, pure flattening does not affect
significantly $\Lx$ and $\Tx$,
while a major reduction in $\Lx$ and $\Tx$ is obtained for the isotropic rotators.

The situation is quite different for the families with low mass progenitors
($\se=200~\kms$). These are the only cases where a transition to a global wind
can be induced by a change of shape or by rotation, in accordance with the energetic
analysis of CP96. This is especially true for the less concentrated Einasto models.
In these global wind cases, $\Lx$ drops to very low values, due
to the very low ISM density, and $\Tx$ keeps larger than expected from the trend defined
by non-wind models (see Sect.~3.3), due to the reduced cooling, and to thermalization
of the meridional motions (see Sects.~\ref{sec:Lx} and \ref{sec:Tx} for a detailed discussion).
The sensitivity of the flow phase for low-mass models near the transition to the outflow is shown
for example by the EO$7^{200}_{\mathrm{VD}}$ model with the NFW halo, that experiences
two quite distinct evolutionary phases (Fig.~\ref{fig5}).
At the beginning, a significant equatorial degassing is apparent, coincident
with the strong heating in that region. As time increases, the velocity field
in the outflow region decreases and gas cooling becomes more and more important
outside the V-shaped region around the equator.
However, after $\simeq9$ Gyr, the secular increase of the specific heating, coupled
with the shallow potential well, induces again higher and higher velocities.
The gas temperature increases again while $\Lx$ decreases.
The associated EO$7^{200}_{\mathrm{IS}}$ model is in a
permanent wind phase from the beginning, thus showing the additional effect of
rotation in flattened, low-mass galaxies. The differences between the EO$7^{200}_{\mathrm{VD}}$ and EO$7^{200}_{\mathrm{IS}}$
models are quantified by the associated values of the global quantities at the
end of the simulation (see Table~\ref{tab:NFW}): $\Mhot = 0.66\times
\tento{9}~\Msun$ and
$0.24\times\tento{9}~\Msun$ in the VD and IS cases, respectively, where $\Mhot$ is the ISM mass having $T>\tento{6}~K$.
Little accretion at the centre is present in the VD but not in the IS, and this shows how
different AGN activity may be expected in rotating vs non rotating galaxies, also
at low galaxy masses.
\subsection{The X-ray ISM luminosity $\Lx$}
\label{sec:Lx}
We now move to describe the properties of $\Lx$ for the whole set of galaxy
models, as they would be observed at an age of 13~$\gyr$.
The results are summarized in Fig.~\ref{fig6}, where the top panels refer to
the NFW set and the bottom panels to the Einasto set. 
$\Lx$ is shown versus 3 different galaxy properties, i.e., $\se$
(left panels), $\Mhot$ (central panels),
and $\Lb$ (right panels).
Remarkably, the range of $\Lx$ values spanned by the models
matches the observed one (see for example the observed $\Lx-L_K$ and $\Lx-\se$ trends
in Figs. 2 and 5 in \citealt{boroson.etal2011}).
The most interesting feature of Fig.~\ref{fig6} is the clear $\Lx$ difference
between flattened rotating models and models of similar $\se$ but velocity dispersion 
supported.
As described in the previous Section, the hydrodynamical simulations show that the under-luminosity
of rotating galaxies with medium to large $\se$ is due to a different flow
evolution driven by the presence of
angular momentum, which prevents the gas from accumulating in the
central regions, leading to the creation of a very hot, low density
atmosphere in the centre, and eventually resulting in a lower total $\Lx$.
\begin{figure}
\includegraphics[width=\linewidth, keepaspectratio]{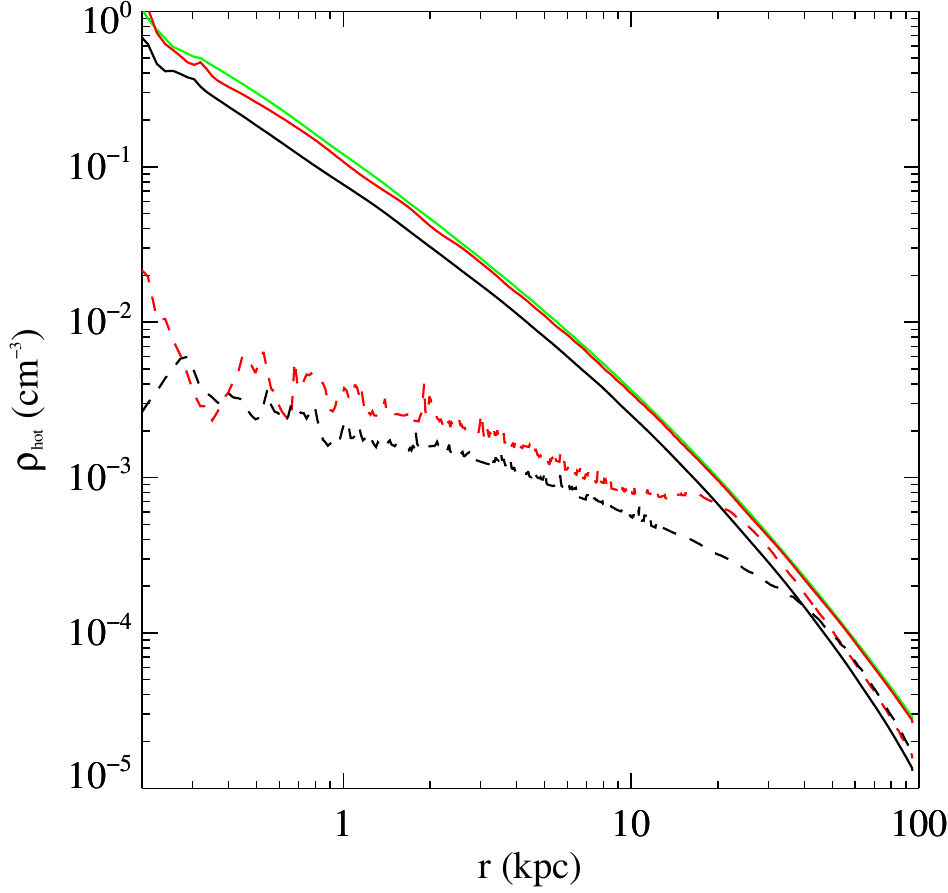}
\caption{Angle-averaged profile of the hot ISM density at $t=13~\gyr$ for 
the same models as in Fig.~\ref{fig2}. Solid lines refer to VD models, dashed lines refer to IS models.}\label{fig7}
\end{figure}
Instead, in VD models the ISM flows directly toward the central galactic regions, where a
steep density profile is created. This difference in the hot gas density distribution is a
major reason for the systematic difference of $\Lx$ (see also Fig.~\ref{fig7}). It nicely explains
the lower $\Lx$ observed for fast rotators than for slow rotators in the ATLAS sample \citep{sarzi.etal.2013}.
ETGs with the lowest $\se$, behave differently (see below).

\begin{figure*}
\includegraphics[width=0.78\linewidth, keepaspectratio]{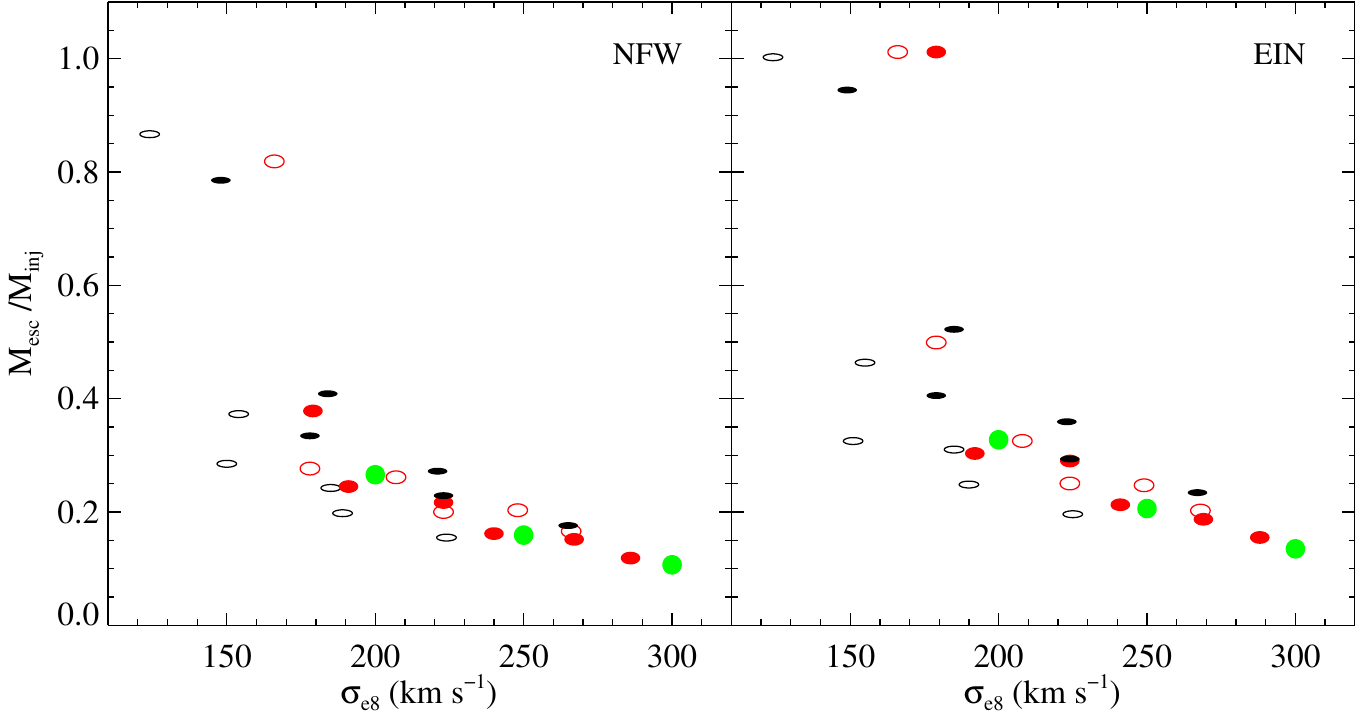}
\caption{Fraction of escaped ISM mass ($\Mesc$) with respect to the
total injected mass ($\Minj$, see Tables~\ref{tab:NFW} and \ref{tab:Ein}) at
$t=13~\gyr$, as a function of $\se$, for the whole NFW and Einasto sets.
The notation for the symbols is the same as in Fig.~\ref{fig6}.}
\label{fig8}
\end{figure*}

In the central panels of Fig.~\ref{fig6} $\Lx$ is plotted against
the hot gas content $\Mhot$; each rotating model is shifted to the left of the corresponding
VD model, thus IS models have also a lower
$\Mhot$ than VD models. This is due to the presence of recurrent 
cooling episodes driven by rotation, that further contribute to the lowering of $\Lx$.
With the exception of the models with the lowest $\Lb$, the systematic
differences in $\Mhot$ are not due to escaping ISM (Fig.~\ref{fig8}).

Finally, the right panels of Fig.~\ref{fig6} show how $\Lx$ on average
increases with the galaxy optical luminosity, however presenting at each $\Lb$
a significant spread in $\Lx$, consistent with observations \citep{boroson.etal2011}. 
At fixed $\Lb$, round progenitors are found at
high $\Lx$, while the dispersion is associated to a mix of flattening and
rotation effects.
At each $\Lb$, $\Lx$ of the VD models is higher than that 
of IS ones by up to a factor of $\simeq 40$. The largest difference 
occurs for the more massive and flatter models,
and it is much larger than the $\Lx$ variation between a
spherical progenitor and its most flattened VD version.
Indeed, $\Lx$ of VD models of identical $\Lb$ with different flattening
lies in a narrow range, with a weak trend for the X-ray luminosity to
increase as the galaxy model gets rounder.
The same behaviour occurs also among IS models with the same
$\Lb$. This indicates that, at fixed $\Lb$ and fixed internal kinematics, $\Lx$ 
is only marginally sensitive to even large variations of the flattening
degree of the stellar component. 

\begin{figure*}
\includegraphics[width=0.8\linewidth, keepaspectratio]{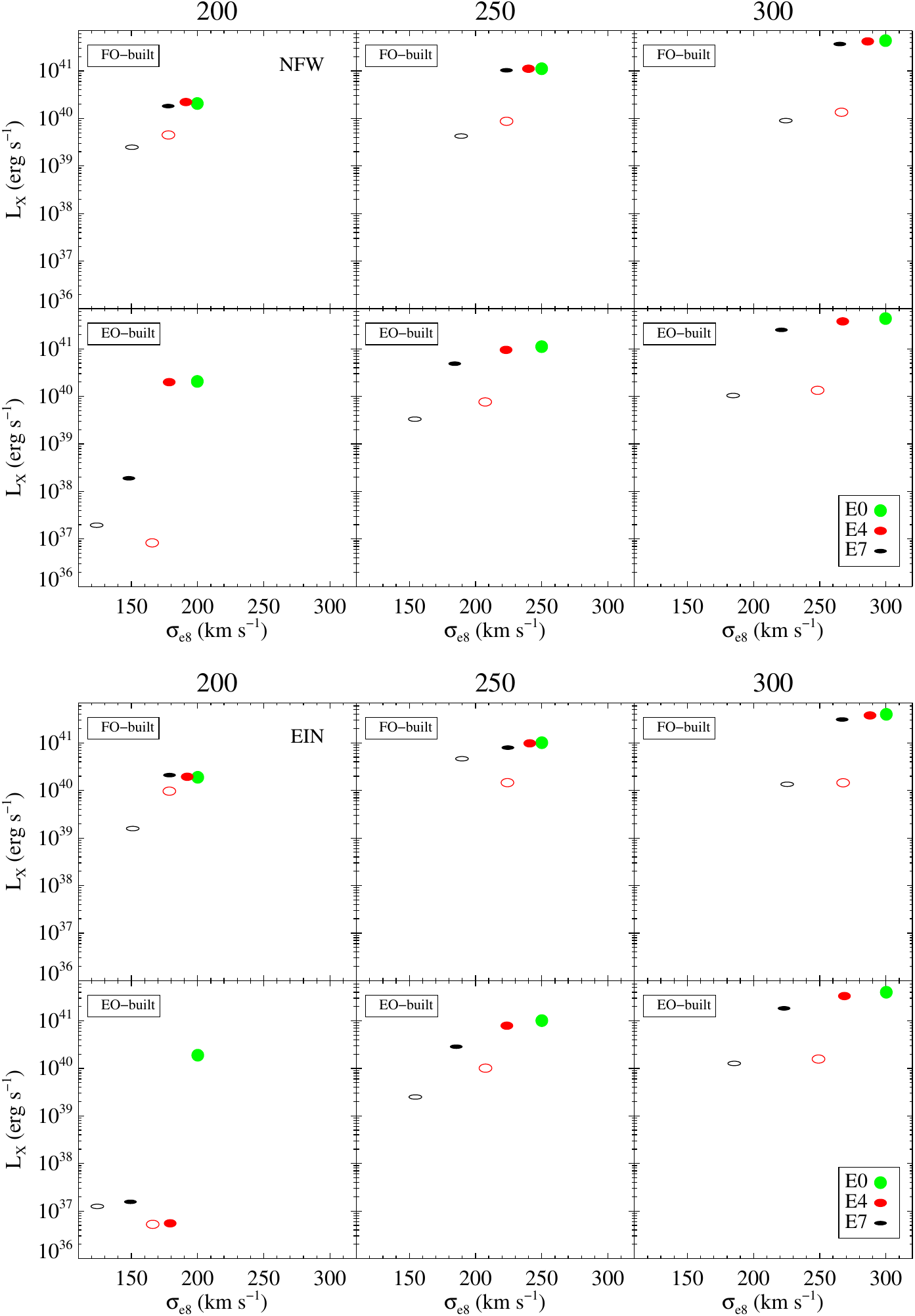}
\caption{ISM X-ray luminosity $\Lx$ in the 0.3--8 keV band at 13 Gyr
for the models in the NFW (top six panels) and in the Einasto
(bottom six panels) sets as a function of $\se$. Different
columns show the results for the families obtained from the
spherical progenitors with $\se=(200, 250, 300)~\kms$, and refer to
model flattened according to the edge-on or face-on procedure.}
\label{fig9}
\end{figure*}

A ``zoom'' on the specific effects of flattening and rotation is given in
Fig.~\ref{fig9}, where we plot, separately for each $\se$, and for FO and EO
built families, the $\Lx$ values of Fig.~\ref{fig6}.
All models in a given column are characterised by the same
optical luminosity. As expected, $\Lx$ of FO sub-families spans a narrower range
of values than that of the EO cases, due the FO flattening procedure in which
the galaxy becomes more concentrated and so outflows are less favoured.
\begin{figure*}
\includegraphics[width=0.8\linewidth, keepaspectratio]{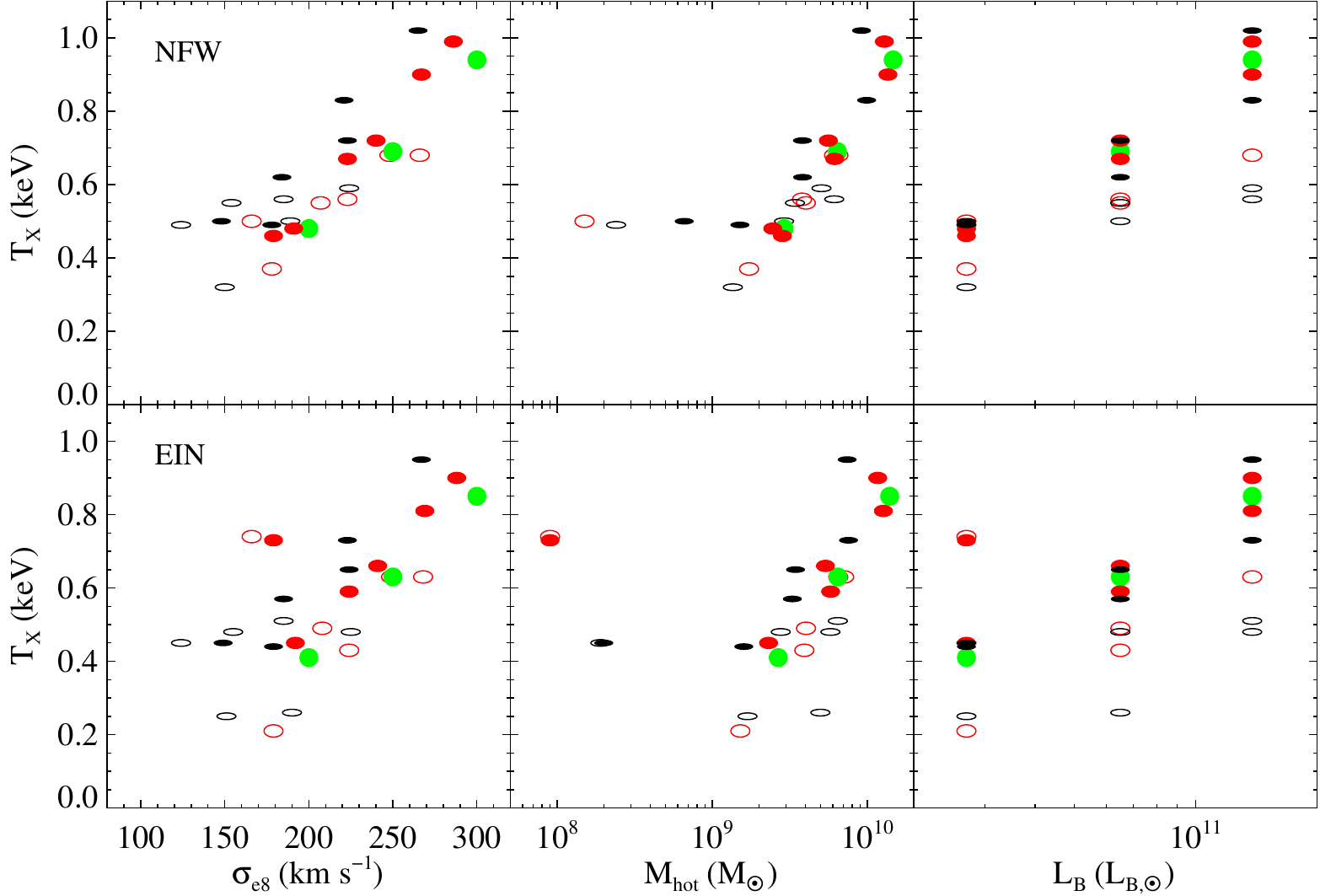}
\caption{ISM emission weighted temperature $\Tx$ in the 0.3--8 keV band at 13
Gyr for all the models in the NFW (top panels) and in the Einasto
(bottom panels) sets as a function of $\se$, of the hot
($T> \tento{6}$~K) ISM mass, and of the galaxy blue optical luminosity.
Symbols and colours are as in Fig.~\ref{fig6}.}
\label{fig10}
\end{figure*}
\begin{figure*}
\includegraphics[width=0.78\linewidth, keepaspectratio]{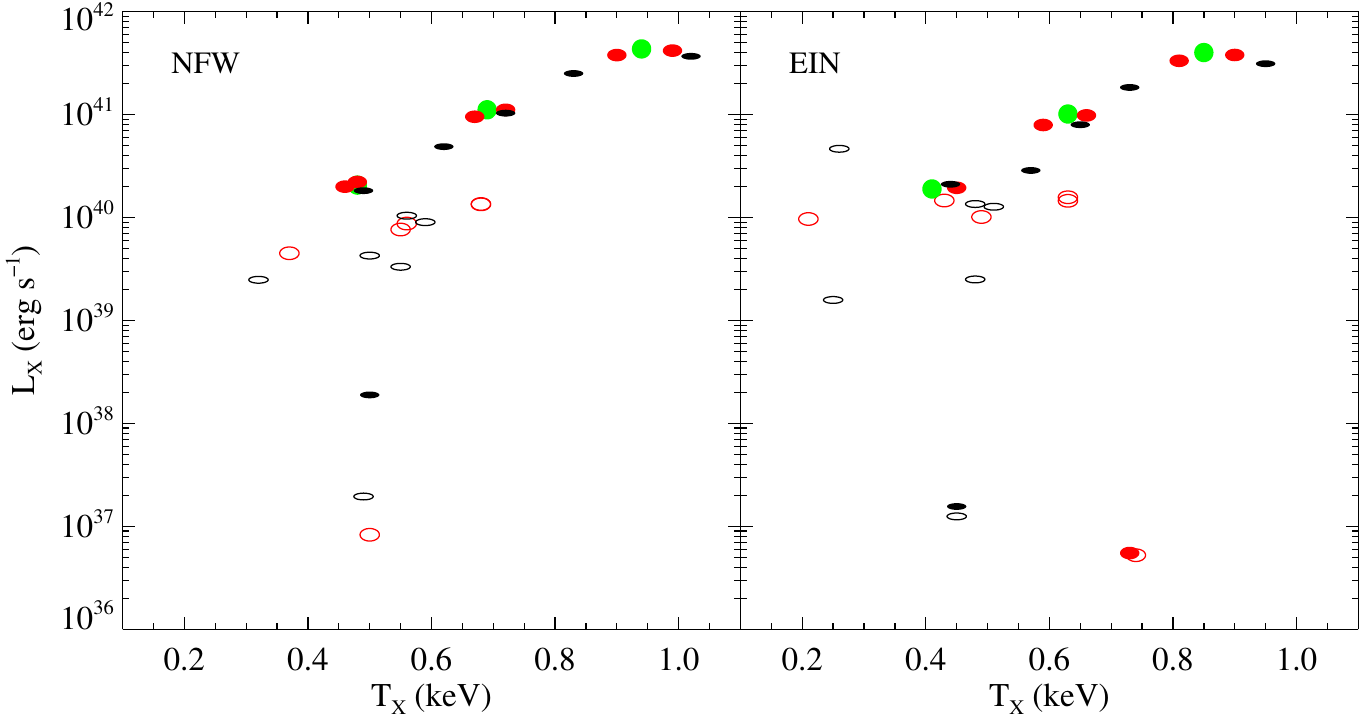}
\caption{X-ray luminosity $\Lx$ with respect to X-ray luminosity weighted temperature
$\Tx$ at $t=13~\gyr$ for the whole NFW and Einasto sets.
The notation for the symbols is the same as in Fig.~\ref{fig6}.}
\label{fig11}
\end{figure*}
\begin{figure*}
\includegraphics[width=0.8\linewidth, keepaspectratio]{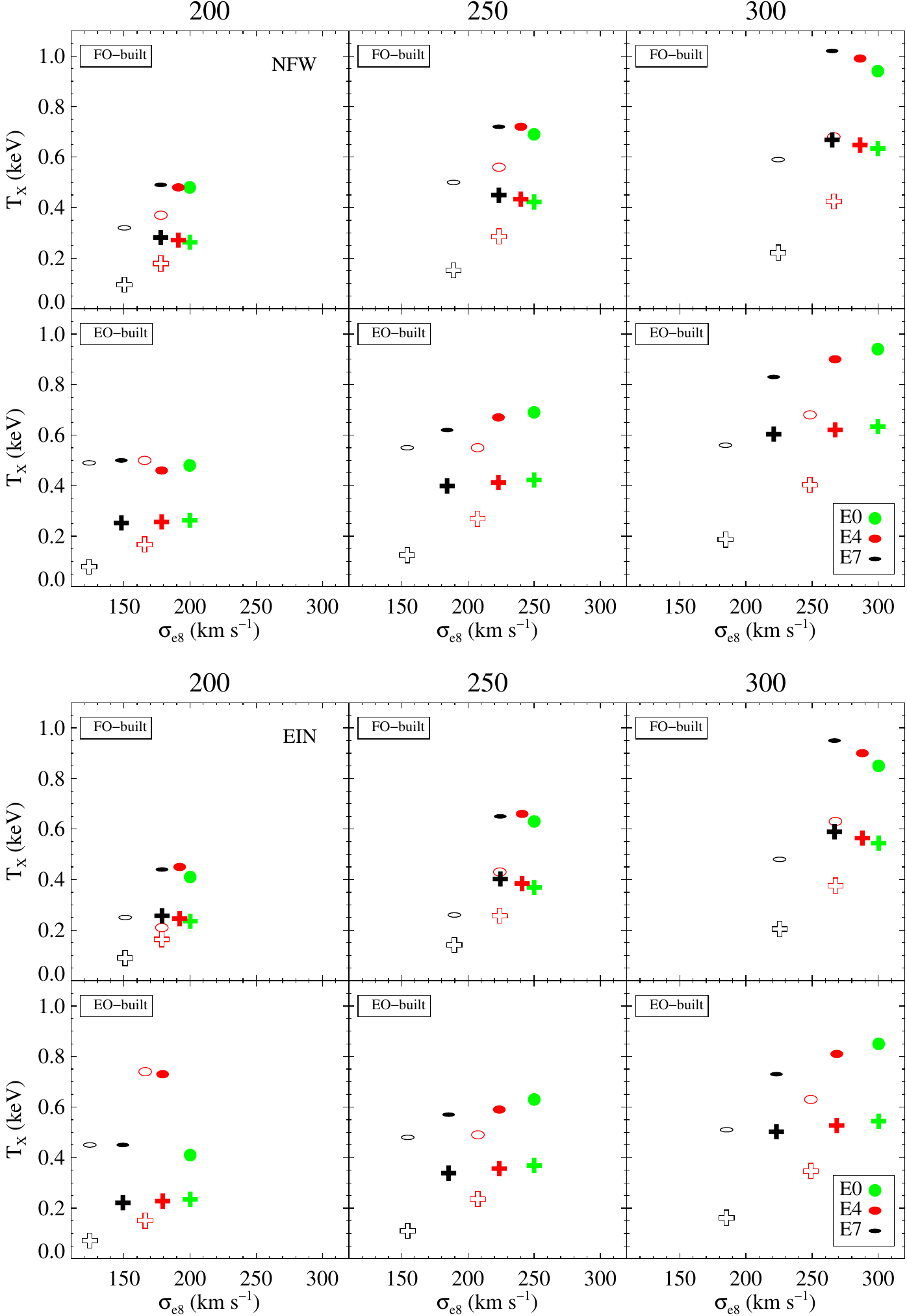}
\caption{The emission weighted ISM temperature $\Tx$ in the 0.3--8
keV band at 13 Gyr for the models in the NFW (top six panels) and in
the Einasto (bottom six panels) sets as a function of 
$\se$. Different columns show the results for the families obtained
from the spherical progenitors with $\se=(200, 250, 300)$ km/s, and
refer to model flattened according to the edge-on or face-on
procedure. The green, red and black colours refer to the E0, E4 and
E7 models respectively (progenitors are in green). Filled symbols
indicate the fully velocity dispersion supported VD models,
while empty symbols indicate the isotropic rotators IS models.
Crosses show the values of $\Tsig$ calculated
according to Eq.~(18). For the relation of $\Tsig$ with $\Tkin$ and $\Tx$,
see the Sect.~2.3.}
\label{fig12}
\end{figure*}

Figure~\ref{fig9} clearly shows how the models with the $\se =200~\kms$ progenitor
behave differently from the rest of the models; this is more evident for the EO flattening,
when the galaxy potential well becomes shallower,
and thus energetic effects of flattening and rotation are larger than for the
FO flattening. For example, the EO$7^{200}_{\mathrm{VD}}$ model drops to low $\Lx$, at variance with
the FO$7^{200}_{\mathrm{VD}}$ model; this drop happens also for the Einasto EO$4^{200}_{\mathrm{VD}}$ model. This sharp $\Lx$
difference is due to the fact that flattening produces a flow transition
to a global wind, in accordance with the CP96
analysis, as described in Sect.~3.1.4. In the NFW case, a further reduction in
$\Lx$ is attained when introducing rotation in the EO$7^{200}_{\mathrm{IS}}$ model, again in
accordance with CP96 and P13, where thermalization of ordered motions does not
take place. Note how a transition to a very low $\Lx$
value is also obtained for the NFW EO$4^{200}_{\mathrm{IS}}$ model, just by adding rotation. These
findings point out the high sensitivity of the flow phase to
(even small) changes in the mass profile (e.g., flattening or mass
concentration) and in the stellar kinematics (e.g., rotation) at low galactic masses,
for which then it is difficult to predict systematic trends in $\Lx$.
We stress that the VD and IS models in
each panel are characterized,
by construction, by the same gravitational potential, so that the
difference in $\Lx$ is only due to galactic
rotation.
\subsection{The X-ray emission weighted temperature $\Tx$} 
\label{sec:Tx}
The second important diagnostic explored is the 0.3--8 keV luminosity
weighted ISM temperature $\Tx$. The distribution of the $\Tx$ values for
the whole set of models at the end of the simulations is given in Fig.~\ref{fig10}, as a function of
$\se$, $\Mhot$ and $\Lb$.

In general $\Tx$ increases with $\se$, a natural consequence of the
deeper potential well associated with larger $\se$. This leads to
faster stellar (random and ordered) velocities, with the consequent
larger energy input from thermalization of the stellar motions.
In addition, in a deeper potential the hot gas is retained at
a larger $\Tx$. The temperature range spanned by the models agrees
well with that of real ETGs, and the observed trend of $\Tx$ with
$\se$ is reproduced (e.g., see Fig.~6 in \citealt{boroson.etal2011},
who measured $\Tx$ of the pure gaseous component for a
sample of 30 ETGs). At high $\se$, the observed $\Tx$ values span a
narrower range than in our models, likely because the models include
very flat and highly rotating ETGs that are missing in the observed
sample. Interestingly, instead, the low-$\se$ end of the observed $\Tx-\se$ relation 
shows an increase of dispersion in the $\Tx$ values, 
and a hint for a flattening of the relation 
with respect to the trend shown at larger $\se$. These features are
shown also by our models: at low $\se$ the trend of $\Tx$ flattens for
NFW models, and the scatter around it increases considerably for the Einasto
models. This is explained as the temperature counterpart of the $\Lx$
behaviour at low $\se$ in Fig.~\ref{fig6}: the transition to global
winds in flattened and rotating low-mass galaxies leads to a reduction
in $\Lx$ and an increase of $\Tx$ with respect to the trend defined by more massive ETGs,
or ETGs of similar mass but not in wind. The change in the relationship is
due to the thermalization of the
resulting meridional flows (while the thermalization of galaxy
rotation remains negligible), and to the lower cooling (see Sect.~3.1.4).
For example, the EO$4^{200}_{\mathrm{VD}}$ and
EO$4^{200}_{\mathrm{IS}}$ models in the Einasto set, have high
$\Tx$ as a consequence of the transition to the wind phase. 

The middle panels of Fig.~\ref{fig10} show the $\Tx$ distribution as a function of $\Mhot$. In the
NFW set, there is a sequence of $\Tx$ values clearly visible at $\Mhot > 2\times10^9\Msun$,
with VD models hotter than the corresponding IS models. However,
the three models with the smallest amount of hot ISM ($\Mhot < 10^9\Msun$)
have higher temperatures than one would expect extrapolating the $\Tx$ sequence to
very low values of $\Mhot$,
as a consequence of the transition to the wind phase.
A change in the trend is even more visible
in the low mass Einasto models, where the stronger tendency to
establish a global wind leads to an \textit{increase} of $\Tx$ at very low
$\Mhot$, reaching values even higher than in VD models with large X-ray haloes.
In conclusion, at medium-high $\se$, $\Tx$ of VD
models tends to remain above that of rotating models; at low $\se$, in
addition to the cooler branch of rotating models, another hotter branch of IS
and VD models appears, made by models in wind.

Finally, the right panels of Fig.~\ref{fig10} show again how
$\Tx$ of IS models is systematically lower with respect to that of VD
ones of same $\Lb$, with the exception of those in the wind phase. As for $\Lx$,
$\Tx$ of VD models is dominated by the dense central luminous regions.
In IS models, instead, the central region is hotter than in VD models, but
it is also at a lower density, so that its contribution
to $\Tx$ is marginal, and $\Tx$ is more affected by colder ($T \simeq 2 \times
\tento{6}~K$) gas located in the outer regions.
Thus, the main reason of the lower $\Tx$ in IS models of medium-high mass is not galaxy shape,
but the importance of galaxy rotation, that drives the hydrodynamical evolution (Sect.~3.1.3).
From the Jeans equations, the more a galaxy is flat, the more
it can be rotating; thus the E7 IS models are cooler than their
VD counterparts, and by a larger amount than for the
analogous E4 pair, due to the stronger rotation in the E7 models. 

The trend of $\Lx$ with $\Tx$ for all models is shown in the lower panels of Fig.~\ref{fig11}. 
Also in this figure the models behaviour is strikingly similar to that observed in the \citet{boroson.etal2011} sample, where
a narrow correlation at high $\Tx\gtrsim 0.5$ keV is broken into an almost vertical band of
$\Lx$ values spanning a large range (from $10^{38}$ to few $10^{41}$ erg s$^{-1}$) 
for $kT$ covering a small range (from 0.2 to 0.5 keV). This trend in the models is explained as the product of the 
effects described above, resulting in a high sensitivity of the flow phase to small variations in the galaxy structure at the
lowest galaxy masses, that on average also have $\Tx<0.5$ keV.

In analogy with Fig.~\ref{fig9}, in Fig.~\ref{fig12} we show the distribution of $\Tx$ of all
models, as a function of $\se$ and of the flattening procedure. The additional
symbols (crosses) represent $\Tsig$ (see Eq.~18), thus they give the temperatures associated with
the thermalization of all stellar velocities for
VD models (solid crosses), and only to the random part of the stellar velocities for IS
models (empty crosses). The values of
$\Tsig$ depend only on the galaxy structure, and do not contain contributions from gas
cooling and SNIa heating.
The simulations show that the values of $\Tkin$ (Eq.~19) are
almost coincident with those of $\Tsig$ in the medium-high $\se$ models (i.e., models in a slow inflow where
$\gth$ is very small, see Tables~\ref{tab:NFW} and \ref{tab:Ein}).
The low-$\se$ wind models, instead, have $\Tkin>\Tsig$, and the temperature difference is due to
thermalization of the strong meridional motions developed in the wind phase 
($\Tkin\simeq\Tsig+\Tm$, while $\Tphi$ remains very small).
In Fig.~\ref{fig12} it is even more apparent than in Fig.~\ref{fig10} how
VD models are in general hotter than their rotating
counterparts, due to the above discussed hydrodynamical effects. 
In addition, the $\Tx$
difference between VD and IS models increases with galaxy flattening,
and it is larger for the more massive and FO-built models, and decreases
for smaller and EO-built models.
Exceptions are found in the low-$\se$ EO-built models, as
a consequence of the transition to global wind induced by flattening and
rotation. 

Two interesting considerations can be made by comparing $\Tx$ resulting from
the simulations with the temperatures $\Tkin$ and $\Tsig$ associated with
the thermalization of stellar motions. The first is that $\Tx$ of all models is
higher than $\Tkin$ and $\Tsig$, as somewhat
expected due to the additional heating contributions (e.g., from SNIa) to the gas,
and to the relatively small radiative losses (we
recall that $\Tx$ is computed from the hot, low-density gas only). The second
consideration is that, notwithstanding the missing SNIa heating and cooling
terms in $\Tsig$, the \textit{trend} of $\Tx$ with galaxy flattening
and rotation is the same as that of $\Tsig$ for all models, with the exception of the global
wind, low-$\se$ models. Thus $\Tsig$, except for wind cases, is a good proxy for $\Tkin$,
and a robust indicator of the trend of $\Tx$ with galaxy properties (shape and internal kinematics).
As a final comment, we note that, in general, at fixed $\se$, Einasto
models tend to be slightly colder than the NFW models, both in $\Tx$ and
$\Tsig$, due to the different dark matter profile.

\section{Discussion and conclusions} 
\label{sec:concl} 
\begin{figure*} 
\includegraphics[width=\linewidth, keepaspectratio]{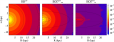}
\caption{Edge-on 0.3--8 keV surface brightness of the ISM ($\Sigmax$) at 13 Gyr,
 for E0$^{250}$, EO7$^{250}_{\mathrm{VD}}$ and EO7$^{250}_{\mathrm{IS}}$ models,
respectively; the
 brightness values on the colour-bar are given in erg s$^{-1}$
 cm$^{-2}$. Superimposed are the isophotes ($\Sigma_\star$) obtained by projecting the
 galaxy stellar density distribution, with the innermost contour corresponding to $10^4$
 $M_{\odot}$ pc$^{-2}$, and decreasing by a
 factor of ten on each subsequent contour going outwards.
Note that the $\Sigmax$ map of the EO7$^{250}_{\mathrm{VD}}$ model shows a 
round shape and a luminous core very similar to the E0$^{250}$ map, 
whereas EO7$^{250}_{\mathrm{IS}}$ map presents a boxy shape and a low-luminosity core.}
\label{fig13}
\end{figure*}
%
In this paper, in a follow-up of a series of preliminary studies, we
performed a large suite of high-resolution 2D
hydrodynamical simulations, to study the effects of galaxy shape and
stellar kinematics on the evolution of the X-ray
emitting gaseous haloes of ETGs. Realistic
galaxy models are built with a Jeans
code, that allows for a full generality in the choice of axisymmetric galaxy shape
and of the stellar and dark matter profiles, that can be tailored to
reproduce observational constraints. The dynamical structure of the
models obeys the implicit assumption of a 2-integrals phase-space
distribution function. Stellar motions in the azimuthal
direction are split among velocity dispersion and ordered rotation
by using the \citet{satoh1980} decomposition. 
In particular, we explored two extreme kinematical configurations,
the fully velocity dispersion supported system (VD) and the isotropic rotator (IS), in order to encompass
all the possible behaviours occurring in nature. Of course,
the VD configuration applies only to a minor fraction of the flat galaxy population
(e.g., \citealt{Emsellem.etal2011}).
Moreover, IS models approximate only to some extent the dynamical structure of flat and fast rotating galaxies, since
the latter are more generally characterized by a varying degree of anisotropy in the meridional plane with intrinsic flattening
\citep{Cappellari.etal2007}.
The source of gas is provided by secular evolution
of the stellar population (stellar winds from ageing stars and
SNIa ejecta). Heating terms account for SNIa events and thermalization of
stellar motions.

The main focus of this work is the explanation of long-standing and more 
recently observed trends of $\Lx$ and $\Tx$ with
galaxy shape and rotation (as well as, of course, with fundamental galaxy
properties as stellar velocity dispersion and optical luminosity).
Evidences from previous exploratory theoretical (CP96, P13) and
numerical works (DC98, N14) seem to point toward a cooperation of
flattening and rotation in establishing the final X-ray luminosity and temperature
of the ISM. However which of the two is the driving parameter, and what
is the involved physical {\it mechanism}, had not been clarified yet.
From the present investigation, we conclude that more than one
physical effect is
at play, and that the relative importance of flattening and rotation changes
as a function of galaxy mass. We summarize the results
discussing first the X-ray luminosity and then the emission-weighted
ISM temperature.

1) In low mass galaxy models with a progenitor hosting a global wind, the effects of flattening and
rotation are just to make the wind stronger, and all systems are found at
the lowest values of $\Lx$.

2) In case of galaxies energetically near to the onset of a galactic wind,
i.e., for ETGs with $\se\approx200\,\kms$, flattening and
rotation contribute significantly to induce a wind, in agreement with the
energetic expectations discussed in CP96, with the consequent sharp decrease of $\Lx$. 
The transition to a global wind is favoured respectively by the facts that
flattening can reduce the depth of the potential well, and that
in rotating systems the ISM and the stellar component almost corotate; this reduces (in absolute value)
the effective potential experienced by the ISM.

3) In models with $\se > 200~\kms$, galaxy shape variations, in absence of rotation,
have only a minor impact on
the values of $\Lx$, in the sense that fully velocity dispersion supported
flattened models have $\Lx$ similar to or just lower than that of their
spherical progenitors.

4) In flat galaxies with $\se > 200~\kms$, rotation reduces significantly $\Lx$.
Not only the thermalization parameter is low and 
part of the heating due to stellar motions is missing with respect to
the corresponding VD model, but rotation
acts also on the hydrodynamics of the gas flow: conservation of
angular momentum of the ISM injected at large radii favours gas cooling
through the formation of rotating discs of cold gas, reducing the amount of hot gas
in the central regions and then $\Lx$. The effects of angular momentum
are clearly visible in Fig.~\ref{fig13}, where we
show the edge-on projected X-ray surface brightness maps. In conclusion, galaxy
flattening has an important, though \textit{indirect} effect for
medium-to-high mass galaxies, in the sense that only flattened systems can host
significant rotation of the stellar component.

5) The luminosity evolution and the 
luminosity values at the end of the simulations are similar for the
NFW or Einasto dark matter haloes (at fixed stellar structure and similar 
values of the dark matter halo mass).

The main results concerning the emission-weighted temperature $\Tx$ can be
summarized as follows:

6) As for $\Lx$, also for $\Tx$ the response to a variation of shape and
internal kinematics is different for low and high mass galaxies.
$\Tx$ does not change appreciably adding flattening and rotation to low mass progenitors that are in the
global wind phase.
Due to their low density and high meridional velocities, global winds are generally hotter than what expected by
extrapolation of the $\Tx$ of more massive systems. As described at point 2) above,
adding flattening and rotation
to ETGs energetically near to host a global wind
leads to a transition to a wind phase, with the consequent increase of $\Tx$.

7) In the medium-high mass galaxies a change of shape produces small changes in $\Tx$.
Adding rotation, instead, results in a much lower $\Tx$. This is because angular momentum conservation
leads to the formation of a massive centrifugally supported cold disc and to a lower density of the hot ISM
in the central regions above and below the equatorial plane,
with respect to VD models. Then, the external, and colder, regions weight more
in the computation of $\Tx$.

8) Overall, for medium-high mass galaxies, $\Tx$ increases with galaxy mass, independently of the specific
dark matter halo profile. In general, in the Einasto haloes the hot gas is
systematically cooler and with a larger scatter in $\Tx$, than in the 
NFW dark matter haloes of comparable mass.

9) In rotating models the ISM almost corotates with the stars, and so there is a
corresponding reduction of the thermalization of the galaxy streaming velocity.
At the same time the rotating ISM is less bound, due to the centrifugal support.
With the exception of low mass galaxies in the wind phase, $\Tsig$ (the temperature
associated with the thermalization of the stellar velocity dispersion) is a good
proxy for $\Tkin$, the true thermalization temperature of stellar motions, as computed
from the simulations; for wind models instead $\Tkin>\Tsig$. In general $\Tx>\Tkin$, but the qualitative
dependence of $\Tx$ on galaxy mass and shape in no wind galaxies is very well reproduced by that of $\Tsig$,
a quantity that can be computed without resorting to numerical simulations.

A few important physical phenomena are still missing from the simulations.
First, it is obvious that in rotationally supported models the massive
and rotating cold discs are natural places for star formation. For observational
studies it would be interesting to estimate age and mass of the new
stars. From the point of view of the present investigation, the formation of stars, by reducing the
amount of cold gas in the equatorial plane, could in
principle also modify the evolution of the ISM. We performed a
few tests where we activated star formation at a rate $\dot{\rho}_*$, following \citet{Ciotti.Ostriker.2012} and
references therein:
\begin{equation}
\dot{\rho}_* = \eta \dfrac{\rho}{\tform},\qquad  \tform=\mathrm{max}(\tcool,\tdyn),
\end{equation}
where $0.01\leqslant \eta \leqslant 0.1$, $\tcool = E/\emissivity$ (Eq.~9), and $\tdyn\propto
1/\sqrt{G\rho}$. In the test simulations for the model FO7$^{300}_{\mathrm{IS}}$ of the NFW set,
star formation peaks at the first $3-4$ Gyr with a rate of $\approx$ 10 $\Msun$ yr$^{-1}$,
and the final mass in the new stellar disc is of the order of few $\times\tento{10}~\Msun$;
most of the gas disc is consumed by star formation.
No significant effects are produced on the ISM luminosity and temperature, so that from this point of view
the present results can be considered robust. However, in the test simulations star formation was activated in a
``passive'' fashion, i.e., only gas subtraction from the computational
domain was considered, so that the injection of mass, momentum, and energy due
to the evolution of the new stars was not included; these effects will be considered in a future
work.
An interesting link between our finding of an ubiquitous formation of a cold disc in rotating systems and
observed galaxy properties is given by the fact that, among ETGs, it is only in fast rotators that some degree of
star formation is observed \citep{Davis.etal2011,Young.etal2011,sarzi.etal.2013}.

A second aspect missing here is the self-gravity of the gaseous cold
disc. It is expected that self-gravity acts not only to promote star formation, but also to develop non
axisymmetric instabilities, that lead to non-conservation of
angular momentum of the gas. Phenomenologically, the effects of self-gravity can be viewed as a
``gravitational viscosity'' \citep[e.g.][]{bertin.lodato.2001}, that favours accretion
of cold gas toward the centre. Such a gas flow toward the centre is
of great importance for feedback effects from a central massive black hole in rotating
galaxies \citep{Novak.etal2011,Gan.etal2014}.
In a complementary exploration, we are currently studying the interplay between the ISM of
a rotating galaxy on large scale and the feedback effects from the central black hole.
\section*{Acknowledgements}
We thank the anonymous referee for constructive comments.
L.C. acknowledges Giuseppe Bertin, Jerry and Eve Ostriker, and James Stone for useful discussions.
L.C. and S.P. were supported by the MIUR grant PRIN 2010-2011, project
`The Chemical and Dynamical Evolution of the Milky Way and Local Group
Galaxies', prot. 2010LY5N2T.
\bibliographystyle{mn2e}
\bibliography{citations.bib}
\appendix
\section{Tables}
\renewcommand\arraystretch{1.4}
 \begin{table*}
 \caption{Simulations results for the NFW set at $t=13$ Gyr.}
 \begin{tabular}{cccccccccccccccccccccccc}
\toprule
 name                    &   $\Minj$  &     $\Mesc$&  $\masstot$ &     $\Mhot$ &    $\Lx$                    &   $\Tx$ &        $\Lsn$ &  $\Tkin$ & $\Tsig$ &  $\Tv$   & $\Tm$ &  $\gth$  & $\gta$    \\
                         &$(10^9\Msun)$&$(10^9\Msun)$&$(10^9\Msun)$&$(10^9\Msun)$& $(10^{40}~\ergs)$         & (keV)   &$(10^{40}~\ergs)$&  (keV) & (keV)   & (keV)    & (keV) &          &           \\ 
 (1)                     &   (2)      &     (3)    &  (4)        &     (5)     &    (6)                      &   (7)   &          (8)  &  (9)     & (10)    &  (11)    & (12)  &   (13)   & (14)      \\
 \midrule                                                                                                                                                                                            
 \midrule                                                                                                                                                                                            
E0$^{200}$               &    12.0    &        3.2 &       8.9  &       2.91  &   2.06                      &    0.48 &           10.2 &    0.26  &    0.26 & 1.7E-3 &  1.7E-3 &   --     &    --     \\
 \midrule                                                                                                                                                                                                  
EO4$^{200}_{\mathrm{IS}}$&    11.9    &        9.8 &       2.4  &       0.15  &  8.29E-4                    &    0.50 &           10.2 &    0.40  &    0.17 & 0.24   &  0.21   &     2.62 &  0.25     \\
EO4$^{200}_{\mathrm{VD}}$&    11.9    &        4.5 &       7.5  &       2.84  &   1.99                      &    0.46 &           10.2 &    0.26  &    0.26 & 1.5E-3 &  1.5E-3 &   --     &    --     \\
EO7$^{200}_{\mathrm{IS}}$&    11.9    &       10.3 &       1.8  &       0.24  &  1.95E-3                    &    0.49 &           10.2 &    0.29  &    0.08 & 0.22   &  0.18   &     1.24 &  0.22     \\
EO7$^{200}_{\mathrm{VD}}$&    11.9    &        9.4 &       2.8  &       0.66  &  1.89E-2                    &    0.50 &           10.2 &    0.33  &    0.25 & 7.8E-2 &  7.8E-2 &   --     &    --     \\
FO4$^{200}_{\mathrm{IS}}$&    12.0    &        3.3 &       8.8  &       1.73  &   0.45                      &    0.37 &           10.3 &    0.21  &    0.18 & 2.9E-2 &  9.0E-3 &     0.31 &  0.21     \\
FO4$^{200}_{\mathrm{VD}}$&    12.0    &        2.9 &       9.2  &       2.46  &   2.21                      &    0.48 &           10.3 &    0.27  &    0.27 & 3.7E-3 &  3.7E-3 &   --     &    --     \\
FO7$^{200}_{\mathrm{IS}}$&    12.0    &        3.4 &       8.7  &       1.36  &   0.25                      &    0.32 &           10.2 &    0.12  &    0.09 & 2.9E-2 &  1.1E-2 &     0.15 &  0.10     \\
FO7$^{200}_{\mathrm{VD}}$&    12.0    &        4.0 &       8.1  &       1.51  &   1.82                      &    0.49 &           10.2 &    0.28  &    0.28 & 2.0E-3 &  2.0E-3 &   --     &    --     \\
 \midrule                                                                                                                                                                                                  
 \midrule                                                                                                                                                                                                  
E0$^{250}$               &    32.2    &        5.1 &      27.5  &       6.43  &   11.1                      &    0.69 &           24.3 &    0.42  &    0.42 & 1.2E-3 &  1.2E-3 &   --     &    --     \\
 \midrule                                                                                                                                                                                                 
EO4$^{250}_{\mathrm{IS}}$&    31.7    &        8.3 &      23.8  &       4.02  &   0.76                      &    0.55 &           23.9 &    0.29  &    0.27 & 1.8E-2 &  3.1E-3 &     0.13 &  0.10     \\
EO4$^{250}_{\mathrm{VD}}$&    31.7    &        6.9 &      25.2  &       6.17  &   9.50                      &    0.67 &           23.9 &    0.42  &    0.41 & 1.5E-3 &  1.5E-3 &   --     &    --     \\
EO7$^{250}_{\mathrm{IS}}$&    30.6    &       11.4 &      19.7  &       3.42  &   0.33                      &    0.55 &           23.1 &    0.18  &    0.13 & 5.5E-2 &  4.9E-3 &     0.20 &  0.18     \\
EO7$^{250}_{\mathrm{VD}}$&    30.6    &       12.5 &      18.7  &       3.83  &   4.87                      &    0.62 &           23.1 &    0.41  &    0.40 & 1.8E-3 &  1.8E-3 &   --     &    --     \\
FO4$^{250}_{\mathrm{IS}}$&    32.2    &        6.5 &      26.2  &       3.80  &   0.87                      &    0.56 &           24.3 &    0.30  &    0.29 & 1.9E-2 &  3.9E-3 &     0.13 &  0.10     \\
FO4$^{250}_{\mathrm{VD}}$&    32.2    &        5.2 &      27.4  &       5.62  &   11.1                      &    0.72 &           24.3 &    0.43  &    0.43 & 1.6E-3 &  1.6E-3 &   --     &    --     \\
FO7$^{250}_{\mathrm{IS}}$&    32.2    &        6.4 &      26.0  &       2.91  &   0.43                      &    0.50 &           24.2 &    0.19  &    0.15 & 3.7E-2 &  7.9E-3 &     0.13 &  0.10     \\
FO7$^{250}_{\mathrm{VD}}$&    32.2    &        7.4 &      25.3  &       3.82  &   10.3                      &    0.72 &           24.2 &    0.45  &    0.45 & 1.7E-3 &  1.7E-3 &   --     &    --     \\
 \midrule                                                                                                                                                                                                 
 \midrule                                                                                                                                                                                                 
E0$^{300}$               &    71.3    &        7.6 &      64.7  &      14.70  &   43.3                      &    0.94 &           49.1 &    0.65  &    0.65 & 1.0E-3 &  1.0E-3 &   --     &    --     \\
 \midrule                                                                                                                                                                                                 
EO4$^{300}_{\mathrm{IS}}$&    69.4    &       14.1 &      56.0  &       6.52  &   1.34                      &    0.68 &           47.7 &    0.43  &    0.42 & 1.5E-2 &  1.9E-3 &     0.07 &  0.06     \\
EO4$^{300}_{\mathrm{VD}}$&    69.4    &       10.5 &      59.9  &      13.61  &   37.7                      &    0.90 &           47.7 &    0.63  &    0.63 & 1.2E-3 &  1.2E-3 &   --     &    --     \\
EO7$^{300}_{\mathrm{IS}}$&    65.5    &       15.9 &      50.2  &       6.13  &   1.04                      &    0.56 &           45.0 &    0.23  &    0.20 & 2.3E-2 &  1.5E-3 &     0.06 &  0.05     \\
EO7$^{300}_{\mathrm{VD}}$&    65.5    &       17.8 &      48.8  &       9.92  &   25.0                      &    0.83 &           45.0 &    0.62  &    0.62 & 1.2E-3 &  1.2E-3 &   --     &    --     \\
FO4$^{300}_{\mathrm{IS}}$&    71.8    &       11.9 &      60.9  &       6.11  &   1.35                      &    0.68 &           49.4 &    0.45  &    0.43 & 1.7E-2 &  4.0E-3 &     0.08 &  0.06     \\
FO4$^{300}_{\mathrm{VD}}$&    71.8    &        8.5 &      64.3  &      12.91  &   41.7                      &    0.99 &           49.4 &    0.65  &    0.65 & 1.5E-3 &  1.5E-3 &   --     &    --     \\ 
FO7$^{300}_{\mathrm{IS}}$&    71.9    &       11.1 &      61.8  &       5.07  &   0.90                      &    0.59 &           49.5 &    0.25  &    0.23 & 2.7E-2 &  4.0E-3 &     0.06 &  0.05     \\
FO7$^{300}_{\mathrm{VD}}$&    71.9    &       12.7 &      60.4  &       9.19  &   36.7                      &    1.02 &           49.5 &    0.67  &    0.67 & 1.6E-3 &  1.6E-3 &   --     &    --     \\
\bottomrule
\end{tabular}
 \flushleft
Notes: (1) Name of the model. $(2)-(3)$ Total ISM mass injected into and escaped from the numerical grid, respectively.
Differences in $\Minj$ for models of same $\Lb$ are accounted for different sampling of $\rhostar$ over the numerical grid.
(4) Total ISM mass retained within the galaxy at the end of the simulation. $(5)-(7)$ ISM mass with $T>10^{6}$ K, ISM X-ray luminosity
in the 0.3--8 keV band, and ISM X-ray emission weighted temperature in the same band, at the end of the simulation.
(8) SNIa heating rate at the end of the simulation. $(9)-(12)$ Thermalization temperatures of stellar motions at the end of the simulation,
defined accordingly to Eqs.~(19) and (18). By construction, $\Tkin=\Tsig+\Tv$; for rotating models $\Tv=\gth\Trot$ and
$\Tphi=\Tv-\Tm=\gta\Trot$, while for velocity dispersion supported models $\Tv=\Tm$ (see Sect.~2.3).
$(13)-(14)$ Thermalization parameter as defined in Eq.~(15), and its azimuthal component $\gta=\Lphi/\Lrot$ (see Eq.~14), at the end of the simulation.
\label{tab:NFW}
\end{table*}

\begin{table*}
\caption{Simulations results for the Einasto set at $t=13$ Gyr.}
\begin{tabular}{ccccccccccccccccccccc}
\toprule
 name                &   $\Minj$ &  $\Mesc$  &$\masstot$ &    $\Mhot$&      $\Lx$                       &  $\Tx$  &  $\Lsn$        &       $\Tkin$&       $\Tsig$&  $\Tv$     & $\Tm$ & $\gth$&  $\gta$   \\
                     &$(10^9\Msun)$&$(10^9\Msun)$&$(10^9\Msun)$&$(10^9\Msun)$& $(10^{40}~\ergs)$        & (keV)   &$(10^{40}~\ergs)$&  (keV)      & (keV)        & (keV)      & (keV) &       &           \\ 
  (1)                &   (2)     &     (3)   &  (4)      &     (5)   &    (6)                           &   (7)   &  (8)           &  (9)         & (10)         &  (11)      & (12)  &  (13) & (14)      \\
\midrule                                                                                                                                                                                                  
\midrule                                                                                                                                                                                                  
E0$^{200}$           &    12.0   &     3.9   &     8.2   &      2.67 &     1.89                         &   0.41  &   10.2         &         0.24 &         0.24 &   3.7E-3 &  3.7E-3 & --    &      --   \\
 \midrule                                                                                                                                                                                               
EO4$^{200}_{\rm{IS}}$&    11.9   &    12.1   &     0.1   &      0.09 &    5.25E-4                       &   0.74  &   10.2         &         0.49 &         0.15 &   0.34   &  0.31   &   4.34&    0.28   \\
EO4$^{200}_{\rm{VD}}$&    11.9   &    12.1   &     0.1   &      0.09 &    5.51-4                        &   0.73  &   10.2         &         0.54 &         0.23 &   0.31   &  0.31   & --    &      --   \\
EO7$^{200}_{\rm{IS}}$&    11.9   &    12.0   &     0.2   &      0.19 &    1.25E-3                       &   0.45  &   10.2         &         0.34 &         0.07 &   0.27   &  0.24   &   1.79&    0.14   \\
EO7$^{200}_{\rm{VD}}$&    11.9   &    11.3   &     0.9   &      0.20 &    1.56E-3                       &   0.45  &   10.2         &         0.42 &         0.22 &   0.20   &  0.20   & --    &      --   \\
FO4$^{200}_{\rm{IS}}$&    12.0   &     6.0   &     5.7   &      1.52 &     0.97                         &   0.21  &   10.3         &         0.21 &         0.16 &   5.2E-2 &  2.1E-2 &   0.63&    0.37   \\
FO4$^{200}_{\rm{VD}}$&    12.0   &     3.6   &     8.5   &      2.31 &     1.94                         &   0.45  &   10.3         &         0.25 &         0.25 &   2.1E-3 &  2.1E-3 & --    &      --   \\
FO7$^{200}_{\rm{IS}}$&    12.0   &     3.9   &     8.2   &      1.69 &     0.16                         &   0.25  &   10.2         &         0.14 &         0.09 &   4.8E-2 &  3.0E-2 &   0.28&    0.11   \\
FO7$^{200}_{\rm{VD}}$&    12.0   &     4.9   &     7.3   &      1.60 &     2.10                         &   0.44  &   10.2         &         0.26 &         0.26 &   2.8E-3 &  2.8E-3 & --    &      --   \\
 \midrule                                                                                                                                                                                               
 \midrule                                                                                                                                                                                               
E0$^{250}$           &    32.2   &     6.6   &    26.0   &      6.47 &     10.1                         &   0.63  &   24.3         &         0.37 &         0.37 &   1.4E-3 &  1.4E-3 & --    &      --   \\
 \midrule                                                                                                                                                                                               
EO4$^{250}_{\rm{IS}}$&    31.7   &    10.3   &    21.8   &      4.03 &     1.01                         &   0.49  &   23.9         &         0.26 &         0.24 &   1.8E-2 &  3.4E-3 &   0.15&    0.12   \\
EO4$^{250}_{\rm{VD}}$&    31.7   &     9.2   &    22.9   &      5.81 &     7.90                         &   0.59  &   23.9         &         0.36 &         0.36 &   1.9E-3 &  1.9E-3 & --    &      --   \\
EO7$^{250}_{\rm{IS}}$&    30.6   &    14.2   &    16.9   &      2.77 &     0.25                         &   0.48  &   23.1         &         0.18 &         0.11 &   6.2E-2 &  9.4E-3 &   0.27&    0.26   \\
EO7$^{250}_{\rm{VD}}$&    30.6   &    16.0   &    15.2   &      3.29 &     2.86                         &   0.57  &   23.1         &         0.35 &         0.34 &   3.2E-3 &  3.2E-3 & --    &      --   \\
FO4$^{250}_{\rm{IS}}$&    32.2   &     8.1   &    24.5   &      3.93 &     1.46                         &   0.43  &   24.3         &         0.27 &         0.26 &   1.7E-2 &  3.8E-3 &   0.13&    0.10   \\
FO4$^{250}_{\rm{VD}}$&    32.2   &     6.9   &    25.8   &      5.38 &     9.79                         &   0.66  &   24.3         &         0.38 &         0.38 &   2.0E-3 &  2.0E-3 & --    &      --   \\
FO7$^{250}_{\rm{IS}}$&    32.2   &     8.0   &    24.4   &      4.99 &     4.64                         &   0.26  &   24.2         &         0.17 &         0.14 &   2.9E-2 &  1.1E-2 &   0.11&    0.07   \\
FO7$^{250}_{\rm{VD}}$&    32.2   &     9.5   &    23.3   &      3.45 &     7.95                         &   0.65  &   24.2         &         0.40 &         0.40 &   2.0E-3 &  2.0E-3 & --    &      --   \\
 \midrule                                                                                                                                                                                               
 \midrule                                                                                                                                                                                               
E0$^{300}$           &    71.3   &     9.6   &    62.7   &     13.97 &     39.9                         &   0.85  &   49.1         &         0.56 &         0.56 &   1.0E-3 &  1.0E-3 & --    &      --   \\
 \midrule                                                                                                                                                                                               
EO4$^{300}_{\rm{IS}}$&    69.4   &    17.2   &    53.2   &      7.10 &     1.57                         &   0.63  &   47.7         &         0.38 &         0.36 &   1.5E-2 &  1.6E-3 &   0.09&    0.08   \\
EO4$^{300}_{\rm{VD}}$&    69.4   &    13.0   &    57.5   &     12.71 &     33.2                         &   0.81  &   47.7         &         0.54 &         0.54 &   1.4E-3 &  1.4E-3 & --    &      --   \\
EO7$^{300}_{\rm{IS}}$&    65.5   &    20.3   &    45.6   &      6.47 &     1.27                         &   0.51  &   45.0         &         0.20 &         0.18 &   2.6E-2 &  2.2E-3 &   0.08&    0.07   \\
EO7$^{300}_{\rm{VD}}$&    65.5   &    23.5   &    43.2   &      7.58 &     18.3                         &   0.73  &   45.0         &         0.52 &         0.52 &   1.8E-3 &  1.8E-3 & --    &      --   \\
FO4$^{300}_{\rm{IS}}$&    71.8   &    14.5   &    58.0   &      6.50 &     1.45                         &   0.63  &   49.4         &         0.40 &         0.38 &   1.8E-2 &  3.6E-3 &   0.10&    0.08   \\
FO4$^{300}_{\rm{VD}}$&    71.8   &    11.1   &    61.8   &     11.70 &     37.8                         &   0.90  &   49.4         &         0.57 &         0.57 &   1.7E-3 &  1.7E-3 & --    &      --   \\
FO7$^{300}_{\rm{IS}}$&    71.9   &    14.1   &    56.6   &      5.79 &     1.35                         &   0.48  &   49.5         &         0.23 &         0.21 &   2.5E-2 &  3.5E-3 &   0.07&    0.06   \\
FO7$^{300}_{\rm{VD}}$&    71.9   &    16.9   &    56.3   &      7.43 &     31.1                         &   0.95  &   49.5         &         0.60 &         0.59 &   1.8E-3 &  1.8E-3 & --    &      --   \\
\bottomrule
\end{tabular}
\flushleft
Notes: all quantities are as in Table~\ref{tab:NFW}.
\label{tab:Ein}
\end{table*}
 \renewcommand\arraystretch{1.}
\label{lastpage}
\end{document}